\documentclass[conference,letterpaper]{IEEEtran}

\usepackage[T1]{fontenc}
\usepackage{cite}
\usepackage{amsmath,amssymb,amsfonts}
\usepackage{graphicx}
\usepackage{textcomp}
\usepackage{xcolor}
\usepackage{booktabs}
\usepackage{tabularx}
\usepackage{microtype}
\usepackage[hyphens]{url}
\usepackage[hidelinks,pdfusetitle]{hyperref}

\makeatletter
\g@addto@macro\UrlBreaks{\do\a\do\b\do\c\do\d\do\e\do\f\do\g\do\h\do\i\do\j\do\k\do\l\do\m
  \do\n\do\o\do\p\do\q\do\r\do\s\do\t\do\u\do\v\do\w\do\x\do\y\do\z
  \do\0\do\1\do\2\do\3\do\4\do\5\do\6\do\7\do\8\do\9}
\makeatother

\newcommand{\adaptshift}{AdaptShift}
\providecommand{\evalnum}[1]{\ifcsname eval@#1\endcsname\csname eval@#1\endcsname\else\textcolor{blue}{\textbf{XX}}\fi}
\expandafter\def\csname eval@dense-saving-min\endcsname{34.1}
\expandafter\def\csname eval@dense-saving-max\endcsname{39.9}
\expandafter\def\csname eval@dense-fixed-saving-ratio-min\endcsname{1.11}
\expandafter\def\csname eval@dense-fixed-saving-ratio-max\endcsname{2.40}
\expandafter\def\csname eval@dense-thrott-saving-ratio-min\endcsname{1.12}
\expandafter\def\csname eval@dense-thrott-saving-ratio-max\endcsname{1.91}
\expandafter\def\csname eval@dense-tafs-ttft-min\endcsname{0.00}
\expandafter\def\csname eval@dense-tafs-ttft-max\endcsname{2.25}
\expandafter\def\csname eval@dense-tafs-tpot-min\endcsname{0.02}
\expandafter\def\csname eval@dense-tafs-tpot-max\endcsname{2.75}
\expandafter\def\csname eval@h20-14-saving\endcsname{39.9}
\expandafter\def\csname eval@h20-14-tpot\endcsname{0.15}
\expandafter\def\csname eval@h20-14-goodput\endcsname{2.346}
\expandafter\def\csname eval@h20-32-saving\endcsname{34.1}
\expandafter\def\csname eval@h20-32-tpot\endcsname{0.60}
\expandafter\def\csname eval@h20-32-goodput\endcsname{1.170}
\expandafter\def\csname eval@gb-14-saving\endcsname{37.1}
\expandafter\def\csname eval@gb-14-tpot\endcsname{0.02}
\expandafter\def\csname eval@gb-14-goodput\endcsname{4.660}
\expandafter\def\csname eval@gb-32-saving\endcsname{34.1}
\expandafter\def\csname eval@gb-32-tpot\endcsname{2.68}
\expandafter\def\csname eval@gb-32-goodput\endcsname{2.200}
\expandafter\def\csname eval@lf-saving\endcsname{24.50}
\expandafter\def\csname eval@lf-tpot\endcsname{1.50}
\expandafter\def\csname eval@lf-ttft\endcsname{0.20}
\expandafter\def\csname eval@lf-goodput\endcsname{3.428}
\expandafter\def\csname eval@busy-saving\endcsname{21.67}
\expandafter\def\csname eval@busy-tpot\endcsname{0.05}
\expandafter\def\csname eval@busy-ttft\endcsname{0.00}
\expandafter\def\csname eval@busy-goodput\endcsname{4.674}
\expandafter\def\csname eval@noadm-saving\endcsname{32.43}
\expandafter\def\csname eval@noadm-tpot\endcsname{29.05}
\expandafter\def\csname eval@noadm-ttft\endcsname{1.05}
\expandafter\def\csname eval@noadm-goodput\endcsname{3.272}
\expandafter\def\csname eval@tafs-saving\endcsname{28.95}
\expandafter\def\csname eval@tafs-tpot\endcsname{1.75}
\expandafter\def\csname eval@tafs-ttft\endcsname{0.00}
\expandafter\def\csname eval@tafs-goodput\endcsname{4.561}
\expandafter\def\csname eval@pd-def-saving\endcsname{0.00}
\expandafter\def\csname eval@pd-def-tpot\endcsname{0.00}
\expandafter\def\csname eval@pd-def-ttft\endcsname{0.00}
\expandafter\def\csname eval@pd-def-goodput\endcsname{2.355}
\expandafter\def\csname eval@pd-def-energy-kj\endcsname{1235.3}
\expandafter\def\csname eval@pd-def-runs\endcsname{1}
\expandafter\def\csname eval@pd-fixed-saving\endcsname{21.36}
\expandafter\def\csname eval@pd-fixed-tpot\endcsname{0.00}
\expandafter\def\csname eval@pd-fixed-ttft\endcsname{0.00}
\expandafter\def\csname eval@pd-fixed-goodput\endcsname{2.354}
\expandafter\def\csname eval@pd-fixed-energy-kj\endcsname{971.4}
\expandafter\def\csname eval@pd-fixed-runs\endcsname{1}
\expandafter\def\csname eval@pd-green-saving\endcsname{28.24}
\expandafter\def\csname eval@pd-green-tpot\endcsname{0.00}
\expandafter\def\csname eval@pd-green-ttft\endcsname{0.00}
\expandafter\def\csname eval@pd-green-goodput\endcsname{2.345}
\expandafter\def\csname eval@pd-green-energy-kj\endcsname{886.4}
\expandafter\def\csname eval@pd-green-runs\endcsname{1}
\expandafter\def\csname eval@pd-tafs-saving\endcsname{37.33}
\expandafter\def\csname eval@pd-tafs-tpot\endcsname{0.00}
\expandafter\def\csname eval@pd-tafs-ttft\endcsname{0.00}
\expandafter\def\csname eval@pd-tafs-goodput\endcsname{2.351}
\expandafter\def\csname eval@pd-tafs-energy-kj\endcsname{774.2}
\expandafter\def\csname eval@pd-tafs-runs\endcsname{1}
\expandafter\def\csname eval@moe-def-saving\endcsname{0.00}
\expandafter\def\csname eval@moe-def-tpot\endcsname{0.35}
\expandafter\def\csname eval@moe-def-ttft\endcsname{0.00}
\expandafter\def\csname eval@moe-def-goodput\endcsname{1.173}
\expandafter\def\csname eval@moe-def-energy-kj\endcsname{2313.0}
\expandafter\def\csname eval@moe-def-runs\endcsname{1}
\expandafter\def\csname eval@moe-fixed-saving\endcsname{17.23}
\expandafter\def\csname eval@moe-fixed-tpot\endcsname{0.45}
\expandafter\def\csname eval@moe-fixed-ttft\endcsname{0.00}
\expandafter\def\csname eval@moe-fixed-goodput\endcsname{1.171}
\expandafter\def\csname eval@moe-fixed-energy-kj\endcsname{1914.4}
\expandafter\def\csname eval@moe-fixed-runs\endcsname{1}
\expandafter\def\csname eval@moe-thrott-saving\endcsname{19.53}
\expandafter\def\csname eval@moe-thrott-tpot\endcsname{95.20}
\expandafter\def\csname eval@moe-thrott-ttft\endcsname{0.00}
\expandafter\def\csname eval@moe-thrott-goodput\endcsname{0.056}
\expandafter\def\csname eval@moe-thrott-energy-kj\endcsname{1861.2}
\expandafter\def\csname eval@moe-thrott-runs\endcsname{1}
\expandafter\def\csname eval@moe-thrott05-saving\endcsname{28.00}
\expandafter\def\csname eval@moe-thrott05-tpot\endcsname{100.00}
\expandafter\def\csname eval@moe-thrott05-ttft\endcsname{0.00}
\expandafter\def\csname eval@moe-thrott05-goodput\endcsname{0.000}
\expandafter\def\csname eval@moe-thrott05-energy kj\endcsname{1665.4}
\expandafter\def\csname eval@moe-thrott05-runs\endcsname{1}
\expandafter\def\csname eval@moe-thrott07-saving\endcsname{19.53}
\expandafter\def\csname eval@moe-thrott07-tpot\endcsname{95.20}
\expandafter\def\csname eval@moe-thrott07-ttft\endcsname{0.00}
\expandafter\def\csname eval@moe-thrott07-goodput\endcsname{0.056}
\expandafter\def\csname eval@moe-thrott07-energy-kj\endcsname{1861.2}
\expandafter\def\csname eval@moe-thrott07-runs\endcsname{1}
\expandafter\def\csname eval@moe-thrott09-saving\endcsname{11.59}
\expandafter\def\csname eval@moe-thrott09-tpot\endcsname{96.55}
\expandafter\def\csname eval@moe-thrott09-ttft\endcsname{0.00}
\expandafter\def\csname eval@moe-thrott09-goodput\endcsname{0.040}
\expandafter\def\csname eval@moe-thrott09-energy-kj\endcsname{2044.9}
\expandafter\def\csname eval@moe-thrott09-runs\endcsname{1}
\expandafter\def\csname eval@moe-tafs-saving\endcsname{25.75}
\expandafter\def\csname eval@moe-tafs-tpot\endcsname{2.57}
\expandafter\def\csname eval@moe-tafs-ttft\endcsname{0.00}
\expandafter\def\csname eval@moe-tafs-goodput\endcsname{1.143}
\expandafter\def\csname eval@moe-tafs-energy-kj\endcsname{1717.3}
\expandafter\def\csname eval@moe-tafs-runs\endcsname{3}
\expandafter\def\csname eval@full-conv-def-saving\endcsname{0.0}
\expandafter\def\csname eval@full-conv-def-ttft\endcsname{1.20}
\expandafter\def\csname eval@full-conv-def-tpot\endcsname{0.00}
\expandafter\def\csname eval@full-conv-def-goodput\endcsname{2.765}
\expandafter\def\csname eval@full-conv-def-energy-kj\endcsname{8491.9}
\expandafter\def\csname eval@full-conv-fixed-saving\endcsname{23.5}
\expandafter\def\csname eval@full-conv-fixed-ttft\endcsname{3.33}
\expandafter\def\csname eval@full-conv-fixed-tpot\endcsname{0.64}
\expandafter\def\csname eval@full-conv-fixed-goodput\endcsname{2.746}
\expandafter\def\csname eval@full-conv-fixed-energy-kj\endcsname{6494.1}
\expandafter\def\csname eval@full-conv-thrott-saving\endcsname{30.6}
\expandafter\def\csname eval@full-conv-thrott-ttft\endcsname{0.00}
\expandafter\def\csname eval@full-conv-thrott-tpot\endcsname{3.10}
\expandafter\def\csname eval@full-conv-thrott-goodput\endcsname{2.678}
\expandafter\def\csname eval@full-conv-thrott-energy-kj\endcsname{5891.0}
\expandafter\def\csname eval@full-conv-tafs-saving\endcsname{34.4}
\expandafter\def\csname eval@full-conv-tafs-ttft\endcsname{2.29}
\expandafter\def\csname eval@full-conv-tafs-tpot\endcsname{0.25}
\expandafter\def\csname eval@full-conv-tafs-goodput\endcsname{2.757}
\expandafter\def\csname eval@full-conv-tafs-energy-kj\endcsname{5573.8}
\expandafter\def\csname eval@full-code-def-saving\endcsname{0.0}
\expandafter\def\csname eval@full-code-def-ttft\endcsname{33.46}
\expandafter\def\csname eval@full-code-def-tpot\endcsname{0.22}
\expandafter\def\csname eval@full-code-def-goodput\endcsname{0.852}
\expandafter\def\csname eval@full-code-def-energy-kj\endcsname{4427.9}
\expandafter\def\csname eval@full-code-tafs-saving\endcsname{27.0}
\expandafter\def\csname eval@full-code-tafs-ttft\endcsname{38.32}
\expandafter\def\csname eval@full-code-tafs-tpot\endcsname{8.63}
\expandafter\def\csname eval@full-code-tafs-goodput\endcsname{0.688}
\expandafter\def\csname eval@full-code-tafs-energy-kj\endcsname{3230.3}
\expandafter\def\csname eval@full-flat-def-saving\endcsname{0.0}
\expandafter\def\csname eval@full-flat-def-ttft\endcsname{0.00}
\expandafter\def\csname eval@full-flat-def-tpot\endcsname{0.00}
\expandafter\def\csname eval@full-flat-def-goodput\endcsname{2.566}
\expandafter\def\csname eval@full-flat-def-energy-kj\endcsname{3440.4}
\expandafter\def\csname eval@full-flat-fixed-saving\endcsname{21.3}
\expandafter\def\csname eval@full-flat-fixed-ttft\endcsname{0.00}
\expandafter\def\csname eval@full-flat-fixed-tpot\endcsname{0.84}
\expandafter\def\csname eval@full-flat-fixed-goodput\endcsname{2.545}
\expandafter\def\csname eval@full-flat-fixed-energy-kj\endcsname{2708.7}
\expandafter\def\csname eval@full-flat-thrott-saving\endcsname{22.5}
\expandafter\def\csname eval@full-flat-thrott-ttft\endcsname{0.00}
\expandafter\def\csname eval@full-flat-thrott-tpot\endcsname{3.14}
\expandafter\def\csname eval@full-flat-thrott-goodput\endcsname{2.486}
\expandafter\def\csname eval@full-flat-thrott-energy-kj\endcsname{2665.5}
\expandafter\def\csname eval@full-flat-tafs-saving\endcsname{22.1}
\expandafter\def\csname eval@full-flat-tafs-ttft\endcsname{0.00}
\expandafter\def\csname eval@full-flat-tafs-tpot\endcsname{0.02}
\expandafter\def\csname eval@full-flat-tafs-goodput\endcsname{2.566}
\expandafter\def\csname eval@full-flat-tafs-energy-kj\endcsname{2679.8}
\expandafter\def\csname eval@full-conv-saving\endcsname{34.4}
\expandafter\def\csname eval@full-conv-tpot\endcsname{0.25}
\expandafter\def\csname eval@full-conv-ttft\endcsname{2.29}
\expandafter\def\csname eval@full-conv-goodput\endcsname{2.757}
\expandafter\def\csname eval@full-conv-goodput-loss\endcsname{0.3}
\expandafter\def\csname eval@full-conv-default-ttft\endcsname{1.20}
\expandafter\def\csname eval@full-conv-default-goodput\endcsname{2.765}
\expandafter\def\csname eval@full-code-saving\endcsname{27.0}
\expandafter\def\csname eval@full-code-tpot\endcsname{8.63}
\expandafter\def\csname eval@full-code-ttft\endcsname{38.32}
\expandafter\def\csname eval@full-code-goodput\endcsname{0.688}
\expandafter\def\csname eval@full-code-goodput-loss\endcsname{19.2}
\expandafter\def\csname eval@full-code-default-ttft\endcsname{33.46}
\expandafter\def\csname eval@full-code-default-goodput\endcsname{0.852}
\expandafter\def\csname eval@flat-code-def-saving\endcsname{0.0}
\expandafter\def\csname eval@flat-code-def-tpot\endcsname{0.00}
\expandafter\def\csname eval@flat-code-def-ttft\endcsname{0.00}
\expandafter\def\csname eval@flat-code-def-goodput\endcsname{2.566}
\expandafter\def\csname eval@flat-code-tafs-saving\endcsname{22.1}
\expandafter\def\csname eval@flat-code-tafs-tpot\endcsname{0.02}
\expandafter\def\csname eval@flat-code-tafs-ttft\endcsname{0.00}
\expandafter\def\csname eval@flat-code-tafs-goodput\endcsname{2.566}
\expandafter\def\csname eval@flat-code-fixed-saving\endcsname{21.3}
\expandafter\def\csname eval@flat-code-fixed-tpot\endcsname{0.84}
\expandafter\def\csname eval@flat-code-fixed-ttft\endcsname{0.00}
\expandafter\def\csname eval@flat-code-fixed-goodput\endcsname{2.545}
\expandafter\def\csname eval@flat-code-thrott-saving\endcsname{22.5}
\expandafter\def\csname eval@flat-code-thrott-tpot\endcsname{3.14}
\expandafter\def\csname eval@flat-code-thrott-ttft\endcsname{0.00}
\expandafter\def\csname eval@flat-code-thrott-goodput\endcsname{2.486}
\expandafter\def\csname eval@sd4-efficiency-gain\endcsname{73.3}
\expandafter\def\csname eval@sd4-def-tpj\endcsname{0.280}
\expandafter\def\csname eval@sd4-tafs-tpj\endcsname{0.485}
\expandafter\def\csname eval@sd2-efficiency-gain\endcsname{66.3}
\expandafter\def\csname eval@sd2-def-tpj\endcsname{0.519}
\expandafter\def\csname eval@sd2-tafs-tpj\endcsname{0.862}
\expandafter\def\csname eval@sd1-efficiency-gain\endcsname{41.7}
\expandafter\def\csname eval@sd1-def-tpj\endcsname{1.012}
\expandafter\def\csname eval@sd1-tafs-tpj\endcsname{1.435}
\expandafter\def\csname eval@sd0.75-efficiency-gain\endcsname{14.2}
\expandafter\def\csname eval@sd0.75-def-tpj\endcsname{1.357}
\expandafter\def\csname eval@sd0.75-tafs-tpj\endcsname{1.550}
\expandafter\def\csname eval@sd0.5-efficiency-gain\endcsname{6.9}
\expandafter\def\csname eval@sd0.5-def-tpj\endcsname{1.567}
\expandafter\def\csname eval@sd0.5-tafs-tpj\endcsname{1.674}
\expandafter\def\csname eval@idle-saving-gap\endcsname{7.28}
\expandafter\def\csname eval@noadm-energy-difference\endcsname{4.41}
\expandafter\def\csname eval@lf-extra-saving\endcsname{-4.45}
\expandafter\def\csname eval@lf-saving-gap\endcsname{4.45}
\expandafter\def\csname eval@pd-green-energy-reduction\endcsname{12.7}
\expandafter\def\csname eval@moe-goodput-loss\endcsname{2.6}
\expandafter\def\csname eval@a40-8-tafs-saving\endcsname{39.7}
\expandafter\def\csname eval@a40-8-tafs-tpot\endcsname{0.20}
\expandafter\def\csname eval@a40-8-thrott-saving\endcsname{40.3}
\expandafter\def\csname eval@a40-8-thrott-tpot\endcsname{1.50}
\expandafter\def\csname eval@sd4-tafs-saving\endcsname{42.30}
\expandafter\def\csname eval@sd4-tafs-tpot\endcsname{0.00}
\expandafter\def\csname eval@sd4-tafs-ttft\endcsname{0.00}
\expandafter\def\csname eval@sd4-def-saving\endcsname{0.00}
\expandafter\def\csname eval@sd4-def-tpot\endcsname{0.00}
\expandafter\def\csname eval@sd4-def-ttft\endcsname{0.00}
\expandafter\def\csname eval@sd2-tafs-saving\endcsname{39.85}
\expandafter\def\csname eval@sd2-tafs-tpot\endcsname{0.15}
\expandafter\def\csname eval@sd2-tafs-ttft\endcsname{0.00}
\expandafter\def\csname eval@sd2-def-saving\endcsname{0.00}
\expandafter\def\csname eval@sd2-def-tpot\endcsname{0.00}
\expandafter\def\csname eval@sd2-def-ttft\endcsname{0.00}
\expandafter\def\csname eval@sd1-tafs-saving\endcsname{29.45}
\expandafter\def\csname eval@sd1-tafs-tpot\endcsname{1.75}
\expandafter\def\csname eval@sd1-tafs-ttft\endcsname{0.00}
\expandafter\def\csname eval@sd1-def-saving\endcsname{0.00}
\expandafter\def\csname eval@sd1-def-tpot\endcsname{0.00}
\expandafter\def\csname eval@sd1-def-ttft\endcsname{0.00}
\expandafter\def\csname eval@sd0.75-tafs-saving\endcsname{12.40}
\expandafter\def\csname eval@sd0.75-tafs-tpot\endcsname{5.30}
\expandafter\def\csname eval@sd0.75-tafs-ttft\endcsname{3.10}
\expandafter\def\csname eval@sd0.75-def-saving\endcsname{0.00}
\expandafter\def\csname eval@sd0.75-def-tpot\endcsname{0.75}
\expandafter\def\csname eval@sd0.75-def-ttft\endcsname{0.30}
\expandafter\def\csname eval@sd0.5-tafs-saving\endcsname{6.42}
\expandafter\def\csname eval@sd0.5-tafs-tpot\endcsname{41.90}
\expandafter\def\csname eval@sd0.5-tafs-ttft\endcsname{85.15}
\expandafter\def\csname eval@sd0.5-def-saving\endcsname{0.00}
\expandafter\def\csname eval@sd0.5-def-tpot\endcsname{25.85}
\expandafter\def\csname eval@sd0.5-def-ttft\endcsname{72.60}
\expandafter\def\csname eval@margin0.1-saving\endcsname{33.46}
\expandafter\def\csname eval@margin0.1-tpot\endcsname{11.50}
\expandafter\def\csname eval@margin0.1-ttft\endcsname{0.15}
\expandafter\def\csname eval@margin0.15-saving\endcsname{31.55}
\expandafter\def\csname eval@margin0.15-tpot\endcsname{2.95}
\expandafter\def\csname eval@margin0.15-ttft\endcsname{0.15}
\expandafter\def\csname eval@margin0.25-saving\endcsname{28.93}
\expandafter\def\csname eval@margin0.25-tpot\endcsname{0.35}
\expandafter\def\csname eval@margin0.25-ttft\endcsname{0.00}
\expandafter\def\csname eval@margin0.35-saving\endcsname{28.40}
\expandafter\def\csname eval@margin0.35-tpot\endcsname{0.35}
\expandafter\def\csname eval@margin0.35-ttft\endcsname{0.00}
\expandafter\def\csname eval@eta0.25-saving\endcsname{29.85}
\expandafter\def\csname eval@eta0.25-tpot\endcsname{2.15}
\expandafter\def\csname eval@eta0.25-ttft\endcsname{0.00}
\expandafter\def\csname eval@eta0.5-saving\endcsname{29.45}
\expandafter\def\csname eval@eta0.5-tpot\endcsname{1.75}
\expandafter\def\csname eval@eta0.5-ttft\endcsname{0.00}
\expandafter\def\csname eval@eta0.75-saving\endcsname{28.89}
\expandafter\def\csname eval@eta0.75-tpot\endcsname{1.50}
\expandafter\def\csname eval@eta0.75-ttft\endcsname{0.00}
\expandafter\def\csname eval@eta0.9-saving\endcsname{27.40}
\expandafter\def\csname eval@eta0.9-tpot\endcsname{0.65}
\expandafter\def\csname eval@eta0.9-ttft\endcsname{0.00}
\expandafter\def\csname eval@iter-a40-8b-prefill-mape\endcsname{11.28}
\expandafter\def\csname eval@iter-a40-8b-decode-mape\endcsname{6.77}
\expandafter\def\csname eval@iter-a40-8b-mixed-mape\endcsname{14.61}
\expandafter\def\csname eval@iter-a40-14b-prefill-mape\endcsname{8.70}
\expandafter\def\csname eval@iter-a40-14b-decode-mape\endcsname{3.99}
\expandafter\def\csname eval@iter-a40-14b-mixed-mape\endcsname{10.81}
\expandafter\def\csname eval@iter-h20-14b-prefill-mape\endcsname{5.30}
\expandafter\def\csname eval@iter-h20-14b-decode-mape\endcsname{3.42}
\expandafter\def\csname eval@iter-h20-14b-mixed-mape\endcsname{7.40}
\expandafter\def\csname eval@iter-h20-32b-prefill-mape\endcsname{8.69}
\expandafter\def\csname eval@iter-h20-32b-decode-mape\endcsname{7.25}
\expandafter\def\csname eval@iter-h20-32b-mixed-mape\endcsname{16.30}
\expandafter\def\csname eval@iter-B200-14b-prefill-mape\endcsname{3.25}
\expandafter\def\csname eval@iter-B200-14b-decode-mape\endcsname{5.42}
\expandafter\def\csname eval@iter-B200-14b-mixed-mape\endcsname{9.20}
\expandafter\def\csname eval@iter-B200-32b-prefill-mape\endcsname{3.71}
\expandafter\def\csname eval@iter-B200-32b-decode-mape\endcsname{6.20}
\expandafter\def\csname eval@iter-B200-32b-mixed-mape\endcsname{12.72}
\expandafter\def\csname eval@iter-B200-flash-prefill-mape\endcsname{3.37}
\expandafter\def\csname eval@iter-B200-flash-decode-mape\endcsname{3.44}
\expandafter\def\csname eval@iter-B200-flash-mixed-mape\endcsname{7.60}
\expandafter\def\csname eval@runtime-B200-14b-mean-ms\endcsname{0.288}
\expandafter\def\csname eval@runtime-B200-14b-p50-ms\endcsname{0.254}
\expandafter\def\csname eval@runtime-B200-14b-max-ms\endcsname{1.36}
\expandafter\def\csname eval@runtime-B200-14b-iter-mean-ms\endcsname{15.47}
\expandafter\def\csname eval@runtime-B200-14b-reported-pct\endcsname{2.15}
\expandafter\def\csname eval@runtime-B200-14b-ratio-of-means-pct\endcsname{1.86}
\expandafter\def\csname eval@runtime-B200-32b-mean-ms\endcsname{0.318}
\expandafter\def\csname eval@runtime-B200-32b-p50-ms\endcsname{0.274}
\expandafter\def\csname eval@runtime-B200-32b-max-ms\endcsname{7.17}
\expandafter\def\csname eval@runtime-B200-32b-iter-mean-ms\endcsname{31.29}
\expandafter\def\csname eval@runtime-B200-32b-reported-pct\endcsname{1.09}
\expandafter\def\csname eval@runtime-B200-32b-ratio-of-means-pct\endcsname{1.02}
\expandafter\def\csname eval@runtime-B200-flash-mean-ms\endcsname{1.745}
\expandafter\def\csname eval@runtime-B200-flash-p50-ms\endcsname{1.594}
\expandafter\def\csname eval@runtime-B200-flash-max-ms\endcsname{8.15}
\expandafter\def\csname eval@runtime-B200-flash-iter-mean-ms\endcsname{35.91}
\expandafter\def\csname eval@runtime-B200-flash-reported-pct\endcsname{5.84}
\expandafter\def\csname eval@runtime-B200-flash-ratio-of-means-pct\endcsname{4.86}

\title{\texorpdfstring{\LARGE}{}%
  AdaptShift: Reducing GPU Energy in LLM Serving\texorpdfstring{\\}{ }%
  under Changing Loads and Variable Clock Transition Delays}
\newcommand{\authoremail}[2]{\href{mailto:#1@#2}{\nolinkurl{#1}}}
\author{%
  \IEEEauthorblockN{%
  Yunfan Cui\textsuperscript{1,3}, Kaixuan Zhang\textsuperscript{1},
  Shuhao Zhang\textsuperscript{1}, Jiaqi Shi\textsuperscript{1},
  Shiyou Qian\textsuperscript{1,*}, Luping Wang\textsuperscript{2,*},
  Jian Cao\textsuperscript{1},\\
  Guangtao Xue\textsuperscript{1}, Cheng Huang\textsuperscript{2},
  Guodong Yang\textsuperscript{2}, and
  Liping Zhang\textsuperscript{2}}
  \IEEEauthorblockA{%
  \textsuperscript{1}Shanghai Jiao Tong University, China\qquad
  \textsuperscript{2}Alibaba Group, China\qquad
  \textsuperscript{*}Corresponding authors\\
  \textsuperscript{3}Work done during an internship at Alibaba Group\\
  \{\authoremail{cuiyunfan}{sjtu.edu.cn}, \authoremail{zks1anx}{sjtu.edu.cn},
  \authoremail{zhang-shuhao}{sjtu.edu.cn}, \authoremail{sjiaq1}{sjtu.edu.cn},
  \authoremail{qshiyou}{sjtu.edu.cn}, \authoremail{cao-jian}{sjtu.edu.cn},
  \authoremail{gt_xue}{sjtu.edu.cn}\}@sjtu.edu.cn}
}
\hypersetup{pdfauthor={Yunfan Cui, Kaixuan Zhang, Shuhao Zhang, Jiaqi Shi,
  Shiyou Qian, Luping Wang, Jian Cao, Guangtao Xue, Cheng Huang, Guodong Yang,
  Liping Zhang}}

\begin{document}

\maketitle
\thispagestyle{plain}
\pagestyle{plain}
\begin{abstract}
Growing demand for LLM inference increases GPU energy consumption and serving costs.
The most energy-efficient GPU frequency varies with the serving load, and a frequency change can take effect after the workload has shifted again.
We present \adaptshift{}, an iteration-level GPU frequency controller that selects energy-efficient operating points under latency service-level objectives (SLOs) and accounts for the time needed to reach them.
The Iteration Latency Predictor composes reusable kernel performance models to predict latency across frequencies. The Dual-objective Frequency Selector uses these predictions and GPU power models to choose the lowest-energy frequency within the iteration's latency budget.
The Adjustment Decision Maker combines the expected workload duration with a measured distribution of downshift effective times to estimate whether a downshift will take effect early enough to save energy.
We implement \adaptshift{} in SGLang and evaluate it on three GPU types using two production traces, covering dense models, mixture-of-experts models, colocated serving, and prefill--decode disaggregation.
\adaptshift{} saves up to 40.2\% GPU energy and achieves up to 1.67$\times$ output tokens per joule (TPJ) relative to NVIDIA's default frequency management. Compared with throttLL'eM and GreenLLM, it achieves up to 1.41$\times$ TPJ while reducing time-to-first-token (TTFT) and time-per-output-token (TPOT) SLO violation rates by up to 2.67$\times$ and 22.5$\times$, respectively.

\end{abstract}

\section{Introduction}
\label{sec:introduction}

Large language models (LLMs) drive conversational assistants and coding agents, creating high demand for GPU inference~\cite{yao2023react,yang2024sweagent}.
For requests with 1,000 input and 1,000 output tokens, Jegham et al.\ estimate 1.215~Wh for GPT-4o and 2.165~Wh for DeepSeek-V3, including data-center overhead~\cite{jegham2025hungry}.
An NVIDIA GB200 NVL72 rack integrates 72 Blackwell GPUs with a reported full-rack power density of 120~kW~\cite{nvidiaGB200nvl72,hamilton2025blackwell}.
The International Energy Agency estimates global data center electricity consumption at 485~TWh in 2025 and projects around 950~TWh by 2030~\cite{iea2026keyquestions}.
Improving inference energy efficiency is crucial for reducing operating costs and easing electricity-supply limits on serving infrastructure~\cite{stojkovic2024greenerllms,iea2026poweringai}.
GPU streaming multiprocessor (SM) frequency scaling offers an energy-saving mechanism, but lower frequencies can prolong execution and violate latency service-level objectives (SLOs)~\cite{ge2013dvfs,kakolyris2025throttllem}.

Researchers have proposed model-based frequency selection policies, which fall into two main categories. 
(1) \emph{Lowest-feasible selection} chooses the lowest predicted SLO-feasible frequency, as in throttLL'eM~\cite{kakolyris2025throttllem}, EcoInfer~\cite{hu2026ecoinfer}, and VoltanaLLM~\cite{yu2025voltanallm}; EcoInfer additionally enforces an energy floor.
(2) \emph{Energy-model optimization} uses latency and power estimates to optimize SLO-feasible frequencies, as in GreenLLM's~\cite{liu2026greenllm} and DualScale's prefill policies ~\cite{basit2026dualscale}; DualScale uses lowest-feasible selection for decode.
Additionally, runtime feedback supports frequency control through online updates in VoltanaLLM and adjustments based on measured decode throughput and latency in GreenLLM.

Existing methods face two limitations.
First, lowest-feasible selection may miss the energy-optimal frequency. For example, GreenLLM's decode feedback relies on recent measurements and incremental adjustments, which can delay response to workload changes ~\cite{liu2026greenllm}.
Second, existing methods ignore frequency transition delays. VoltanaLLM treats application delay as negligible ~\cite{yu2025voltanallm}, while DualScale relies on safety margins ~\cite{basit2026dualscale}.
Neither explicitly compares downshift savings against transition costs over the expected workload duration.

Our evaluations reveal two key observations that guide frequency control (Section~\ref{sec:motivation}).
First, the frequency--energy relationship is non-monotonic, and the optimal frequency varies with serving load.
Second, a frequency command affects kernel execution only after a variable \emph{transition effective time}.
These delays shorten downshift saving windows and prolong low-frequency execution during upshifts.
Effective frequency control therefore requires load-adaptive energy selection that accounts for transition time and workload persistence.

We present \adaptshift{}, an iteration-level GPU frequency-scaling framework that combines load-adaptive energy optimization with transition-time awareness under latency SLO constraints. \adaptshift{} consists of three modules.
The Iteration Latency Predictor (ILP) uses kernel performance predictions to estimate iteration latency at candidate frequencies.
The Dual-objective Frequency Selector (DFS) uses these predictions to estimate energy consumption as workloads change and selects the optimal frequency among SLO-feasible candidates.
The Adjustment Decision Maker (ADM) prioritizes upshifts to protect latency SLOs and uses transition effective-time distributions and workload persistence to evaluate downshifts.

We implement \adaptshift{} in SGLang~\cite{zheng2024sglang}.
It supports dense and mixture-of-experts (MoE) models, including prefill (P), decode (D), and mixed iterations, in PD colocated and disaggregated deployments.
We compare \adaptshift{}  with four baselines across six dense Qwen3~\cite{qwen2025qwen3} deployments on NVIDIA A40, H20, and GB200 using two production traces from Microsoft Azure~\cite{azure2023llmtrace} and Alibaba Qwen-Bailian~\cite{qwenbailiantrace}.
Additional Azure-trace experiments cover PD disaggregation and MoE DeepSeek-V4-Flash serving~\cite{deepseek2026v4}.
Across all traces and dense deployments, \adaptshift{} reduces GPU energy by 20.0--40.2\% relative to NVIDIA's default frequency policy, achieving up to 1.67$\times$ output tokens per joule (TPJ).
Compared with throttLL'eM~\cite{kakolyris2025throttllem} and GreenLLM~\cite{liu2026greenllm}, \adaptshift{} achieves up to 1.41$\times$ TPJ and reduces TTFT and TPOT SLO violation rates by up to 2.67$\times$ and 22.5$\times$, respectively.

This paper makes the following three contributions:
\begin{itemize}
  \item \textbf{Two patterns in GPU frequency scaling.}
    We characterize a non-monotonic frequency--energy relationship with load-dependent optima and variable transition effective times that limit downshift savings.

  \item \textbf{A scaling framework that models both patterns.}
    We design \adaptshift{} to model both patterns under latency SLO constraints, selecting optimal frequencies using transition-time distributions and workload persistence.

  \item \textbf{Implementation and large-scale evaluation.}
    We implement \adaptshift{} in SGLang for dense and MoE models, supporting colocated serving and PD disaggregation across three GPU platforms and two production traces.
\end{itemize}

\section{Motivating Observations}
\label{sec:motivation}
\label{sec:background-motivation}

\begin{figure}[t]
    \centering
    \includegraphics[width=\linewidth]{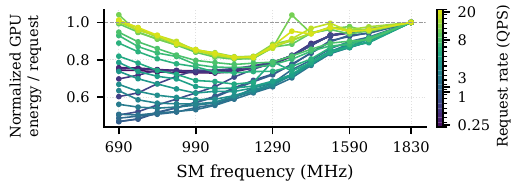}
    \caption{GPU energy per request, normalized to the energy at 1830~MHz for each QPS setting.}
    \label{fig:frequency-energy-qps}
\end{figure}

Our measurements reveal two key observations: the energy-optimal frequency depends on the serving load, and frequency transitions consume a portion of the time available for energy savings. 

\subsection{Observation 1: Energy-Optimal Frequencies Depend on Serving Load.}
\label{sec:frequency-energy-motivation}

Lower GPU power does not necessarily result in reduced total serving energy~\cite{ge2013dvfs,stojkovic2024greenerllms}.
Reducing SM frequency lowers power draw but may prolong execution, thereby extending the duration over which the GPU consumes energy~\cite{kakolyris2025throttllem,liu2026greenllm}.
Conversely, increasing frequency modifies this trade-off: faster execution can offset higher power draw until the energy cost of the increased power exceeds the performance gains~\cite{ge2013dvfs,liu2026greenllm}.
Therefore, an energy-efficient serving strategy involves balancing power and execution time when dynamically adjusting frequencies~\cite{stojkovic2025dynamollm,kakolyris2025throttllem}.

Our evaluations indicate that the serving load alters this balance, even when the model and request length remain consistent.
Figure~\ref{fig:frequency-energy-qps} shows fixed-frequency measurements of Qwen3-14B on H20 with 512 input tokens and 128 output tokens per request.
We vary the request rate (queries per second, QPS), and compare the GPU energy per request across different SM frequencies.

Figure~\ref{fig:frequency-energy-qps} reveals two frequency regions exhibiting different energy characteristics.
Reducing frequency from 1,830~MHz to 1,200~MHz generally lowers energy per request across the measured loads. However, at lower frequencies, the impact depends on the serving load. Among sampled loads exceeding 3~QPS, further frequency reduction eventually increases energy once the frequency drops below the workload-dependent optimum. At sampled loads of 3~QPS or lower, energy remains relatively stable or continues to decrease as frequency is reduced.

\begin{figure}[t]
    \centering
    \includegraphics[width=\linewidth]{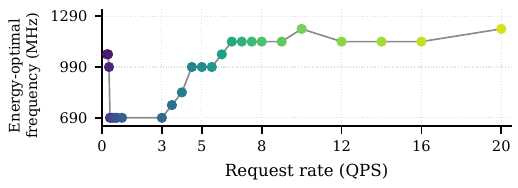}
    \caption{\textbf{The energy-optimal frequency.}}
    \label{fig:frequency-energy-optima}
\end{figure}

Figure~\ref{fig:frequency-energy-optima} illustrates this workload dependence. There is a U-shaped relationship between energy and frequency. Our evaluations show that the \emph{energy-optimal frequency} increases from 690~MHz at 3~QPS to about 1,140~MHz at 8~QPS.
Additionally, adjacent frequencies near the minimum exhibit similar energy, forming an optimal frequency band. 
Similar trends occur on other GPUs.

\noindent\textbf{Implication: proactively adapting frequency selection to the workload.}
An adaptive frequency-scaling framework should compare the energy costs of candidate frequencies as the serving workload fluctuates.
Additionally, interactive serving introduces latency constraints (SLOs) to the selection process~\cite{zhong2024distserve,kakolyris2025throttllem,wu2023fastserve,fu2024serverlessllm}.
This motivates integrating latency and power predictions to select a frequency that improves energy efficiency while staying within the latency budget under dynamic workloads.
In our framework \adaptshift{}, the ILP and DFS modules address this objective (Section~\ref{sec:dual-objective-frequency-selector}).

\subsection{Observation 2: Frequency Transition Delay Limits the Energy-saving Efficiency.}
\label{sec:transition-effective-time-motivation}

Under dynamic workloads, frequency scaling benefits from minimizing adjustment latency~\cite{velicka2025switching}. We define the \emph{transition effective time} as the interval between the issuance of a frequency command and its observable impact on kernel execution~\cite{velicka2025switching}. With continuous batching~\cite{yu2022orca,kwon2023vllm}, request arrivals and departures between forward iterations cause the workload to fluctuate~\cite{kakolyris2025throttllem}. A frequency shift intended for an upcoming iteration may only take effect after the workload has changed, thereby narrowing the opportunity window.

\begin{figure}[t]
    \centering
    \includegraphics[width=0.85\linewidth]{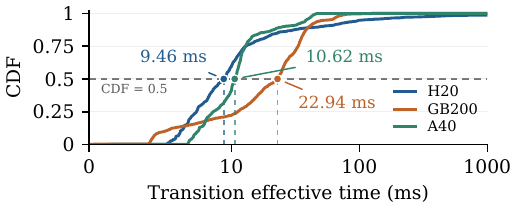}
    \caption{\textbf{Transition effective times exhibit long tails.} 
    The time axis is linear below 10~ms and logarithmic above it.}
    \label{fig:transition-effective-time}
\end{figure}

Figure~\ref{fig:transition-effective-time} shows that transition effective times span tens of milliseconds and vary across trials.
We pool 2,000 downshift trials per GPU on A40, H20, and GB200.
On A40, 40.95\% of trials show a sustained performance response within 10~ms, and all do so within 100~ms.
On H20, 59.5\% of trials show a response within 10~ms, and 94.7\% do so within 100~ms.
The corresponding fractions on GB200 are 22.3\% and 99.65\%, respectively.
A40 and H20 have similar median transition effective times of 10.62 and 9.46~ms, respectively.
GB200's median is 22.94~ms, about 2.4$\times$ H20's.

Our data indicate that transition effective times and decode iterations span comparable timescales.
For Qwen3-14B on an H20 with a batch size of 8 and a KV length of 2,048, reducing frequency from 1,590 to 1,140~MHz increases median forward latency from 14.23 to 17.05~ms, while decreasing energy per forward from 5.80 to 4.69~J.
Combining these profiles with measured transition effective time distributions allows us to estimate the loss in energy savings.
The H20 GPU operates at the current frequency until the transition completes, then switches to the target frequency.
An instantaneous downshift saves 19.1\% compared to a steady 1,590 MHz baseline. Accounting for transition effective time reduces estimated savings to 6.9\% for one forward replay and 14.1\% for five consecutive identical replays.
Given high-frequency replay durations of 14.23 and 71.14~ms, shorter workloads exhibit a greater reduction in potential energy savings.

Using the median to represent transition time also underestimates upshift latency risks.
At a batch size of 64 with the same LLM, GPU, and KV length, median forward latencies at 1,590~MHz and 1,140~MHz are 42.76~ms and 59.26~ms, respectively.
Let $T_h$ and $T_l$ denote the forward latencies at fixed high and low frequencies, respectively.
Suppose we issue an upshift at iteration start and it takes effect after $d<T_l$.
The iteration completes a fraction $d/T_l$ of its work at the low frequency and the remainder at the high frequency, giving $L(d)=d+\left(1-\frac{d}{T_l}\right)T_h.$
Thus, a 50~ms iteration budget requires $d\leq(50\,\mathrm{ms}-T_h)/(1-T_h/T_l)\approx26.00$~ms.
Although the median transition time is 9.46~ms on H20, 25\% of trials exceed this threshold.
Relying on the median suggests the budget is met while overlooking long-tail violations.


\noindent\textbf{Implication: Applying distinct strategies for frequency upshifts and downshifts.}
The two adjustment directions serve different purposes.
When the current frequency fails to meet SLOs, postponing an upshift exacerbates the delay in reaching the target performance. Consequently, the controller prioritizes prompt upshifts.
In contrast, the benefits of a downshift depend on the duration of the lower frequency and transition latency.
A short-lived target or a prolonged transition significantly reduces these gains.
These observations motivate the ADM module of \adaptshift{} in Section~\ref{sec:adjustment-decision-maker}.

\section{\adaptshift{} Framework}
\label{sec:framework}

\begin{figure}[t]
    \centering
    \includegraphics[width=\columnwidth]{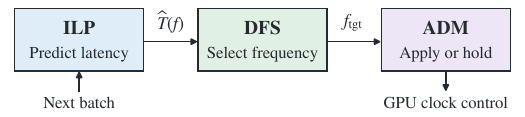}
    \caption{Overall framework of \adaptshift{}.}
    \label{fig:framework}
\end{figure}

\subsection{Basic Idea}
Our objective is to reduce GPU energy consumption by matching frequency (computing power) to dynamic workload while satisfying SLOs.
This matching spans two workload granularities.
At the service level, frequency should match the overall serving load reflected by the request rate~\cite{stojkovic2025dynamollm,kakolyris2025throttllem}.
At the iteration level, frequency should adapt to the computational requirements of each forward pass.
Accounting for both granularities allows frequency control to respond to changes in overall serving load while meeting the execution demands of each iteration.

Several factors complicate this problem, including workload dynamics, frequency-energy nonlinearity, SLO constraints, and adjustment timeliness. In essence, 
maintaining this match requires deciding both when and how to adjust the frequency.
First, changes in serving load and iteration operations call for revisiting frequency decisions, but delayed transitions may make an adjustment take effect after the workload has changed (Section~\ref{sec:transition-effective-time-motivation}).
The timing of an energy-saving adjustment therefore depends on how long the workload is expected to favor the selected frequency.
Second, choosing the target frequency requires balancing power against performance: a lower frequency reduces power but may raise total energy by prolonging execution (Section~\ref{sec:frequency-energy-motivation}).

\textbf{When to adjust.}
To set appropriate adjustment timing, we consider four aspects.
First, continuous batching is a common way to raise GPU utilization in LLM serving~\cite{yu2022orca, kwon2023vllm,duan2024muxserve}.
As new requests join and completed ones leave, batch composition can differ across forward iterations~\cite{yu2022orca,kwon2023vllm}.
Second, during decode, growing KV caches change the amount of attention work and can alter iteration latency even when batch membership is unchanged~\cite{kakolyris2025throttllem,wu2024loongserve}.
Third, 
as co-scheduled requests in the same batch share a GPU frequency, the system cannot assign an independent frequency to each request.
Fourth, transition effective time also limits kernel-level control and can exceed the execution time of individual kernels (Section~\ref{sec:transition-effective-time-motivation}).
\adaptshift{} therefore adopts iteration-level frequency control.

\textbf{How to adjust.}
Accurate frequency adjustments require characterizing the links among frequency, performance and energy. To reduce complexity, \adaptshift{} adopts a two-stage policy: forming a set of SLO-feasible frequencies and selecting a frequency based on the non-monotonic frequency-energy relationship. 
First, to check whether a candidate frequency meets the performance constraints, \adaptshift{} predicts the upcoming iteration's latency at that frequency and compares it with a budget derived from the request-level SLOs.
Second, the U-shaped frequency-energy relationship means that the lowest SLO-feasible frequency need not minimize energy (Section~\ref{sec:frequency-energy-motivation}).
\adaptshift{} therefore combines its iteration latency predictor with power models for execution and idle periods to estimate energy at each SLO-feasible frequency, 
selecting the frequency with the lowest predicted energy.

\subsection{Overall Framework}
\label{sec:overall-framework}

\adaptshift{} coordinates latency prediction, energy-aware frequency selection, and frequency adjustment at the granularity of each forward iteration. As shown in  Figure~\ref{fig:framework}, the framework consists of three components: the \emph{Iteration Latency Predictor (ILP)}, the \emph{Dual-objective Frequency Selector (DFS)}, and the \emph{Adjustment Decision Maker (ADM)}.

ILP predicts the upcoming iteration's latency at different frequencies, providing the performance basis for scaling.
Its design exploits the computational kernels shared across LLMs while accounting for each model's execution structure.
Separating shared kernel behavior from model-specific composition makes latency predictors reusable across LLMs in the same deployment environment. This reuse reduces profiling and retraining when deploying other LLMs within that scope.

DFS treats latency as a feasibility constraint and energy as the optimization objective, a process manifested in two distinct stages. First, DFS links request-level SLOs to iteration-level frequency selection, ensuring local decisions remain aligned with the serving objective. Second, it integrates ILP performance estimates with power estimates to compare energy consumption for SLO-feasible frequencies. This formulation captures the non-monotonic frequency--energy relationship and leverages available latency headroom to select the target frequency with the lowest predicted energy.

ADM decides whether to apply the frequency selected by DFS by evaluating its expected duration and the time required for the transition. This design addresses the mismatch between per-iteration decisions and delayed GPU frequency transitions. A selected frequency may lose its advantage if the workload changes before the transition completes. Consequently, ADM factors in workload persistence and transition-time variability to assess potential energy savings. It prioritizes upshifts to maintain SLOs and allows downshifts only when expected savings justify the transition costs. 
This collaboration enables DFS to select target frequencies each iteration, while ADM regulates the actual GPU adjustments.

\subsection{Iteration Latency Predictor (ILP)}
\label{sec:iteration-latency-predictor}

Frequency selection depends on precise latency predictions to identify SLO-feasible frequencies and compare energy consumption. Generalization, accuracy, and efficiency are vital for ILP. 
First, ILP should generalize across LLMs to limit redundant profiling and training. 
Second, 
prediction errors skew energy estimates. Thus, ILP needs to capture how workload and LLM execution structures influence latency across SM frequencies.
Third, ILP should also keep its total prediction overhead small relative to iteration execution time, since it evaluates candidate frequencies before each iteration.

Different LLMs share computational kernels despite their variations in workload, input shapes, and invocation counts~\cite{agrawal2024vidur,zhang2026pipeweave,aminabadi2022deepspeed}. ILP exploits this by predicting kernel latency based on workload and SM frequency, then composing the predictions according to the iteration's execution structure. This approach captures workload-dependent kernel behavior and model-specific execution structure without locking predictors to a single LLM. 

\subsubsection{Kernel-level Latency Predictors}
\label{sec:kernel-level-latency-predictors}

\paragraph{Across-LLM Generalization} 
Different LLMs share a common set of dominant computational kernels, while their execution workloads mainly differ in kernel types, input shapes, and invocation counts~\cite{agrawal2024vidur,zhang2026pipeweave}.
ILP thus predicts the latency of these kernels, and composes them to obtain iteration latency.
All kernel-level predictors in ILP are implemented as random-forest (RF) regressors~\cite{breiman2001randomforests} and are trained offline using profiled kernel latency samples collected across SM frequencies and workload configurations. 
The kernel-level predictors are trained separately for each GPU type.
They do not directly generalize across GPU types, as kernel latency depends on the GPU architecture and the implementations selected by software libraries~\cite{yu2021habitat,lee2025neusight,shah2024flashattention3}.

\begin{table}[t]
\caption{Features of kernel-level latency predictors.}
\label{tab:kernel-predictors}
\centering
\scriptsize
\begin{tabular}{@{}l@{\hspace{1em}}l@{}}
\toprule
Kernel & Features \\
\midrule
GEMM & $f, M, N, K, \mathrm{FLOPs}$ \\
Prefill attention & $f, B, \ell_{\mathrm{chunk}}, \ell_{\mathrm{prior}}, n_h, n_{\mathrm{kv}}, d_h$ \\
Decode attention & $f, B, \ell_{\mathrm{kv}}, n_h, n_{\mathrm{kv}}, d_h$ \\
Collective comm & $f, s_{\mathrm{bytes}}$ \\
\bottomrule
\end{tabular}
\end{table}

Table~\ref{tab:kernel-predictors} presents the input features of the kernel-level latency predictors.
Every predictor takes the SM frequency $f$ as an input feature.
For GEMM~\cite{nvidia2026cutlass,deepseek2025deepgemm}, $M$, $N$, and $K$ specify the matrix dimensions, and $\mathrm{FLOPs}=2MNK$ gives the floating-point operation count~\cite{nvidia2026cublas}.
For attention, $B$ denotes the batch size; $\ell_{\mathrm{chunk}}$ and $\ell_{\mathrm{prior}}$ denote the current chunk length and the KV-cache length before processing that chunk during prefill, respectively.
During decode, $\ell_{\mathrm{kv}}$ denotes the length of the KV cache that the attention kernel accesses.
The head configuration comprises the number of query heads ($n_h$), the number of key/value heads ($n_{\mathrm{kv}}$), and the dimension of each head ($d_h$)~\cite{vaswani2017attention,shazeer2019mqa,ainslie2023gqa}.
For collective communication, $s_{\mathrm{bytes}}$ denotes message size in bytes~\cite{nvidia2026nccl}.

\begin{table}[t]
\caption{Modeled kernel-time coverage on H20 at 1830 MHz.}
\label{tab:operator-coverage}
\centering\scriptsize
\begin{tabular*}{\columnwidth}{@{\extracolsep{\fill}}lrrrrr@{}}
\toprule
Model & Phase & GEMM + LMhead & Attn. & Comm. & Total \\
\midrule
Qwen3-14B (TP1) & Prefill & 83.5\% & 13.4\% & 0      & 96.9\% \\
               & Decode  & 84.6\% & 12.6\% & 0      & 97.1\% \\
Qwen3-32B (TP4) & Prefill & 64.9\% & 17.8\% & 12.4\% & 95.1\% \\
               & Decode  & 65.7\% & 15.9\% & 14.2\% & 95.7\% \\
\bottomrule
\end{tabular*}
\end{table}

Table~\ref{tab:operator-coverage} shows that the modeled operators account for 95.1--97.1\% of total GPU kernel time across prefill and decode.
The measurements cover Qwen3-14B and Qwen3-32B with tensor parallelism (TP) degrees of 1 and 4, respectively, on H20 GPUs at an SM frequency
of 1,830~MHz.
This coverage indicates that the kernel predictors in ILP capture the dominant components of iteration latency.

\paragraph{Efficiency Optimization}
Although the kernel-level predictors are trained as RF regressors, ILP does not invoke them directly at runtime. 
Instead, the offline profiling stage evaluates each predictor on a predefined grid of SM frequencies and workload configurations and stores the resulting predictions in lookup tables. 
At runtime, ILP queries these tables and interpolates between neighboring entries.
Specifically, inverse-frequency interpolation gives a local approximation $T(f)\approx a/f+b$, combining first-order compute scaling~\cite{guerreiro2017classification,liu2026greenllm} with a frequency-independent offset.
The neighboring table values determine $a$ and $b$, accommodating weaker frequency dependence when memory or communication dominates~\cite{williams2009roofline}.
For continuous workload dimensions, including matrix dimensions, batch size, sequence lengths, and message size, ILP performs linear interpolation in the original feature domain. 
Discrete configuration parameters, such as attention head geometry, are matched directly to the corresponding profiled entries.

The lookup tables are bounded by the profiled workload ranges. 
Inputs below the range are clamped to the nearest boundary, while inputs above the range are treated as out-of-distribution (OOD) and trigger a conservative fallback to the highest candidate frequency.
In CPU microbenchmarks with Qwen3-14B on H20, predicting iteration latency across all 77 candidate frequencies takes an average of $55.1\,\mu\mathrm{s}$ for TP1 and $60.9\,\mu\mathrm{s}$ for TP4.

\subsubsection{Iteration-level Latency Estimation}
\label{sec:iteration-level-latency-estimation}

The kernel-level latency predictions are aggregated according to the execution structure of the upcoming iteration to obtain its latency under each candidate SM frequency. This execution structure depends on the model architecture, parallelization strategy, and backend kernel implementations~\cite{agrawal2024vidur,zhang2026pipeweave}. For example, TP execution can add collective communication, while prefill and decode have different attention workloads and input shapes that may lead the backend to select different implementations~\cite{dao2022flashattention,dao2023flashattention2,ye2025flashinfer}.

For a dense transformer, ILP composes the predicted latency of each transformer layer, the TP communication within that layer, and the LM head. With $L$ layers, one attention-output all-reduce and one MLP-down all-reduce per layer,
the iteration latency is estimated as
\begin{equation}
\begin{aligned}
\widehat{T}_{\mathrm{dense}}(f)
=L\big(&
T_{\mathrm{gemm}}+T_{\mathrm{attn}}+2\,T_{\mathrm{allreduce}}
\big)+T_{\mathrm{head}}.
\end{aligned}
\label{eq:dense-latency}
\end{equation}
The two all-reduce operations arise from the TP execution of the attention output projection and MLP down projection~\cite{shoeybi2019megatron}.
When TP is disabled, the corresponding communication terms are omitted.

For MoEs~\cite{fedus2022switch,jiang2024mixtral}, ILP extends the dense model by replacing the MLP with routed-expert computation and DeepEP~\cite{deepseek2025deepep} communication.
Expert computation depends on active experts and their received tokens~\cite{rajbhandari2022deepspeedmoe,hwang2023tutel,gale2023megablocks,lepikhin2020gshard}.
ILP estimates each rank's expert-token load per layer using cumulative DeepEP recorder counts.
It calculates this by multiplying the ratio of total received expert-tokens to total input-tokens by the upcoming input-token count. This estimate informs iteration-latency predictions at each candidate frequency. While this average captures persistent routing imbalances, it overlooks iteration-level fluctuations.

DeepEP transmits tokens to expert ranks and receives their outputs~\cite{deepseek2025deepep}.
Because expert workloads vary across ranks, ILP utilizes the slowest rank's predicted expert time per layer:
\begin{equation}
T_{\mathrm{EP},l}(f)
=D_l(f)+\max_i E_{l,i}(f)+C_l(f).
\label{eq:moe-ep-stage}
\end{equation}
For layer $l$, $D_l$ and $C_l$ are DeepEP dispatch and combine times, and $E_{l,i}$ is the expert computation time on rank $i$.
ILP predicts these times based on input and received expert-token counts per rank.
The maximum expert time accounts for delays from slower ranks.
At frequency $f$, ILP predicts the total iteration latency across $L$ layers as:
\begin{equation}
\begin{aligned}
\widehat{T}_{\mathrm{MoE}}(f)
={}&\sum_{l=1}^{L}\big[T_{\mathrm{gemm},l}(f)+T_{\mathrm{attn},l}(f)+T_{\mathrm{allreduce},l}(f)\\
&+T_{\mathrm{EP},l}(f)\big]+T_{\mathrm{head}}(f).
\end{aligned}
\label{eq:moe-latency}
\end{equation}

\subsection{Dual-Objective Frequency Selector (DFS)}
\label{sec:dual-objective-frequency-selector}

DFS faces two coupled difficulties when it selects an SM frequency for each iteration. First, request-level TTFT and TPOT include GPU execution together with queuing and scheduling delays~\cite{agrawal2024sarathi,zhong2024distserve,li2023alpaserve}, whereas ILP predicts the GPU execution time of one iteration. DFS therefore needs to convert each applicable request-level SLO into an iteration-level latency budget that allocates slack for non-GPU delays. The conversion follows the iteration type, including prefill, decode, and mixed iterations.
Second, lowering SM frequency reduces instantaneous power but lengthens execution time, so total energy varies non-monotonically with frequency. Consequently, the lowest SLO-feasible frequency may consume more energy than a higher one, while the energy-optimal frequency may violate SLOs.

DFS addresses these challenges in two stages. First, it maps TTFT and TPOT requirements to an iteration-level latency budget and uses ILP predictions to construct the SLO-feasible frequency set. Second, it estimates the energy of each SLO-feasible frequency using predicted execution time and frequency-dependent power estimates, then selects the energy-optimal frequency. The following subsections detail the two stages.

\subsubsection{Forming the SLO-Feasible Frequency Set}
\label{subsec:slo-budget}

DFS first maps request-level SLOs to an effective latency budget for the upcoming iteration. Let $\mathrm{SLO}_{\mathrm{req}}$ denote the phase-specific request-level target. DFS uses TTFT and TPOT requirements for prefill and decode iterations, respectively. A safety margin $m\in[0,1)$ is set to account for ILP prediction error and host-side overhead incurred during an iteration, such as scheduling, batch preparation, and execution-state management. DFS captures their aggregate effect through the margin factor $(1-m)$.

Additionally, higher QPS can cause waiting requests to accumulate when arrivals outpace processing~\cite{sun2024llumnix,zhong2024distserve}.
DFS uses the waiting-queue length to capture this pressure and tightens the iteration latency budget accordingly.
Let $q$ denote the current waiting-queue length, and let $\gamma \geq 0$ control the sensitivity to queue pressure.
DFS computes the effective iteration budget as:
\begin{equation}
\mathrm{SLO}_{\mathrm{ite}}
=
\dfrac{\mathrm{SLO}_{\mathrm{req}}(1-m)}{1+\gamma q}.
\label{eq:slo-budget}
\end{equation}
When $q=0$, no queue-based tightening applies.
For $\gamma>0$, a larger waiting queue tightens the budget to require faster execution.

For chunked prefill, DFS computes Eq.~\ref{eq:slo-budget} for each request and divides the resulting budget by its remaining chunk count, including the current chunk. It uses the smallest resulting budget across prefill requests for the iteration.
For mixed iterations that contain both prefill and decode work, DFS uses the tighter of the two phase-specific effective budgets. 

After deriving $\mathrm{SLO}_{\mathrm{ite}}$, DFS filters the set of candidate SM frequencies $\mathcal{F}$. Specifically, for each candidate frequency $f\in\mathcal{F}$, ILP predicts the GPU execution time of the upcoming iteration, denoted by $\widehat{T}(f)$. DFS retains candidates whose predicted latency satisfies the effective iteration budget:

\begin{equation}
\mathcal{F}_{\mathrm{feas}}
=
\left\{
f\in\mathcal{F}\mid
\widehat{T}(f)\leq \mathrm{SLO}_{\mathrm{ite}}
\right\}.
\label{eq:feasible-set}
\end{equation}

A frequency is SLO-feasible if its predicted GPU execution time satisfies the SLO-derived iteration budget.
DFS compares energy estimates only among the SLO-feasible frequencies in $\mathcal{F}_{\mathrm{feas}}$.

\subsubsection{Frequency-Energy Relationship Modeling}
\label{sec:power-modeling-energy}

\begin{figure}[t]
    \centering
    \includegraphics[width=\columnwidth]{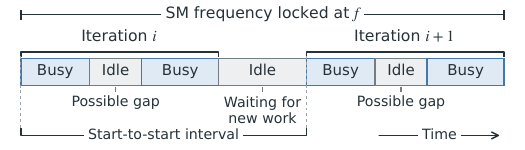}
    \caption{Schematic of GPU busy and idle intervals.}
    \label{fig:busy-idle-window}
\end{figure}

DFS estimates energy consumption in two GPU states:
busy, when the GPU runs kernels, and idle, when no kernels execute~\cite{hong2010integrated}, as shown in Figure~\ref{fig:busy-idle-window}. The GPU maintains a constant SM frequency in both states to avoid frequent frequency shifts. Such shifts can consume energy saved during short idle intervals and delay the return to a performance-sufficient frequency, increasing SLO violation risks.
Reducing SM frequency lowers power but can increase execution time, so the lowest SLO-feasible frequency need not minimize energy.
DFS therefore combines ILP latency estimates with separate busy and locked-idle power models.

\noindent\textbf{Busy Power Modeling.} The dynamic power follows $P_{\mathrm{dyn}}\propto V^2f$~\cite{leng2013gpuwattch}. With voltage increasing approximately linearly with frequency, that is, $V \propto f$, dynamic power scales approximately as $P_{\mathrm{dyn}}\propto f^3$~\cite{kandiah2021accelwattch,wang2025analyticaldvfs}. Following the polynomial form used in GreenLLM~\cite{liu2026greenllm}, we model busy power as
\begin{equation}
P_{\mathrm{busy}}(f)
= a_0+a_1f+a_2f^2+a_3f^3.
\label{eq:busy-power}
\end{equation}
Here, $f$ denotes the normalized SM frequency,\footnote{We divide the SM frequency in MHz by 1000 to reduce scale differences among the polynomial terms during coefficient fitting.} and we fit the coefficients $a_0,\ldots,a_3$ to GPU profiling measurements.

\noindent\textbf{Idle Power Modeling.} Locked-idle power is the GPU power consumption when no kernels execute while the SM frequency remains locked~\cite{liu2026greenllm}.
Figure~\ref{fig:busy-idle-window} shows two types of idle interval: optional gaps within an iteration and waits for new work between iterations.
The start-to-start interval includes both types, during which the GPU consumes energy at the locked SM frequency.
We measure this power at sampled SM frequencies, and DFS estimates $P_{\mathrm{idle}}(f)$ by linearly interpolating these measurements.
DFS uses this model to estimate idle energy during host-side batch preparation and waits for new work.

\noindent\textbf{State-aware Energy Modeling.} Consecutive iterations usually exhibit similar workload characteristics due to temporal locality in continuous batching~\cite{hu2026ecoinfer}. DFS therefore uses the observed inter-iteration interval to estimate the reference window for the upcoming iteration, smoothing timing fluctuations with an exponential moving average.
Specifically, DFS uses the iteration time, measured between consecutive iteration starts, as a reference window for energy estimation.
Let $\delta_i$ denote the latest observed inter-iteration interval available before iteration $i$.
Before iteration $i$, DFS updates the reference window using $\delta_i$:
\begin{equation}
T_{\mathrm{iter},i}
=
(1-\alpha)T_{\mathrm{iter},i-1}
+\alpha\delta_i,
\label{eq:iter-ema}
\end{equation}
where $\alpha\in(0,1]$ controls the weight of the latest observation.
For each frequency $f$, DFS estimates energy as
\begin{equation}
\begin{aligned}
E(f)
={}&P_{\mathrm{busy}}(f)\widehat{T}(f)
+P_{\mathrm{idle}}(f)
\left[
T_{\mathrm{eff}}(f)-\widehat{T}(f)
\right],
\end{aligned}
\label{eq:energy}
\end{equation}
where
$T_{\mathrm{eff}}(f)=
\max(T_{\mathrm{iter}},\widehat{T}(f))$.
When the predicted execution time $\widehat{T}(f)$ exceeds the recent iteration interval $T_{\mathrm{iter}}$, the maximum function extends the reference window to ensure the estimated idle duration remains non-negative. The idle term approximates the non-execution energy within this window.
Lower QPS can increase idle time between iterations, changing the energy trade-off among SLO-feasible frequencies. DFS accounts for this effect by including both execution and idle energy within a reference window estimated from recent iteration-start intervals.

Prefill and decode exhibit different patterns of inter-iteration idle time.
Under saturated decode, the scheduler continuously processes active sequences, resulting in relatively short inter-iteration gaps and making busy energy the dominant contribution.
Prefill iterations, in contrast, involve additional host-side batch preparation and execution-state management, which can introduce longer gaps between GPU executions and increase the contribution of locked-idle energy.
Sparse request arrivals can similarly increase idle time during decode.

\subsubsection{Dual-Objective Frequency Selection}
\label{sec:dual-objective-selection}

DFS selects an SM frequency that satisfies the SLO while minimizing the estimated energy consumption.
Specifically, it selects the energy-optimal frequency among the SLO-feasible frequencies:

\begin{equation}
f_{\mathrm{tgt}} =
\begin{cases}
\displaystyle \arg\min_{f\in\mathcal{F}_{\mathrm{feas}}} E(f),
& \mathcal{F}_{\mathrm{feas}}\neq\varnothing,\\[8pt]
f_{\max},
& \mathcal{F}_{\mathrm{feas}}=\varnothing.
\end{cases}
\label{eq:ftgt}
\end{equation}

Thus, DFS prioritizes the latency constraint before minimizing energy consumption within the SLO-feasible frequency set. 
This process repeats at each iteration, utilizing the current latency and power estimates.

\subsection{Adjustment Decision Maker (ADM)}
\label{sec:adjustment-decision-maker}

DFS selects the frequency with the lowest predicted energy among SLO-feasible options, but this selection assumes execution at the target frequency. A delayed downshift may take effect after the workload changes, reducing its energy benefit and exposing subsequent iterations to insufficient frequency.

To address this, ADM uses distinct policies for frequency upshifts and downshifts. When DFS selects a higher frequency, ADM prioritizes the upshift to mitigate SLO violation risks. For downshifts, ADM evaluates whether the expected energy savings outweigh the associated transition opportunity costs, accounting for workload persistence and uncertainty in transition effective time.

ADM assesses a downshift by considering three primary factors. First, the predicted energy difference between the current and target frequencies reflects the potential savings per iteration. Second, the history of DFS selections allows for an estimation of the duration for which energy-saving opportunities persist. Third, the transition opportunity cost is influenced by both the power difference and the transition duration. ADM characterizes the uncertainty in this duration using the measured transition-latency distribution.

ADM uses these estimates to assess a downshift’s potential energy benefit under its transition-cost model. It calculates the total energy gain over the expected workload horizon and derives the maximum transition duration that this gain can support. ADM then utilizes the empirical transition-latency distribution to estimate the probability of adjustment completion within this timeframe. It executes the downshift only when the projected energy saving is positive and this probability satisfies the required threshold.

\subsubsection{Frequency Transition Effective Time CDF}
\label{sec:effective-time-cdf}

ADM uses offline measurements of frequency-transition latency to construct an empirical cumulative distribution function (CDF).
The transition latency measures the time from issuing a frequency-setting command to the point at which the kernel execution time indicates that the target frequency has taken effect~\cite{velicka2025switching}.

Let $s_1,\ldots,s_N$ denote the observed transition times.
For a duration $t$, the empirical CDF is defined as
\begin{equation}
\widehat{F}(t)
=
\frac{1}{N}
\sum_{i=1}^{N}
\mathbf{1}\!\left(s_i\leq t\right),
\label{eq:empirical-cdf}
\end{equation}
where $\mathbf{1}(\cdot)$ is the indicator function, which returns one if the condition is met and zero otherwise. Consequently, $\widehat{F}(t)$ represents the fraction of observed transitions completed within time $t$.

\subsubsection{Frequency Persistence Estimation}
\label{sec:frequency-persistence}

ADM estimates the persistence of DFS selections as an indicator of how long a downshift remains effective. Section~\ref{sec:frequency-energy-motivation} demonstrates that multiple neighboring frequencies yield similar energy levels near the minimum. This observation supports the aggregation of persistence statistics across proximate frequencies. DFS selections may oscillate between adjacent frequencies as iteration characteristics shift. Tracking each frequency individually would fragment a continuous period of similar target selections into shorter intervals when the frequency changes. ADM consequently groups neighboring targets into bands to track how long DFS selects frequencies in the same band.

The band width $w$ defines the resolution of this approximation. Wider bands decrease run fragmentation but may merge targets with varying latency and energy traits. Narrower bands maintain these distinctions yet render persistence estimates more sensitive to minor changes in DFS selections. Note that bands serve exclusively to aggregate persistence statistics; DFS continues to evaluate individual frequencies, and ADM utilizes the specific current and target frequencies to estimate energy savings and transition costs.

Specifically, ADM partitions the GPU's supported SM frequency range into $K$ contiguous, non-overlapping bands:
\begin{equation}
\mathcal{B}
=
\{\mathcal{B}_1,\mathcal{B}_2,\ldots,\mathcal{B}_K\}.
\label{eq:frequency-bands}
\end{equation}
ADM establishes these boundaries prior to the serving process to ensure that historical persistence estimates remain associated with consistent frequency ranges. For instance, in our H20 experiments, we partition the 690--1830~MHz range into 11 bands. Each of the first ten bands covers 100~MHz, while the final band spans $(1690,1830]$~MHz.

ADM defines a band episode as a sequence of consecutive iterations where the DFS selections remain within the same band. An episode concludes when DFS selects a frequency from a different band. For each band $\mathcal{B}_k$, ADM tracks completed episodes chronologically by index $j$, denoting the length of the $j$-th episode in iterations as $D_{k,j}$. These lengths serve as historical samples reflecting the duration for which DFS selects a specific band. ADM applies an exponential moving average to smooth these samples:
\begin{equation}
\bar{D}_{k,j+1}
=
\beta D_{k,j+1}
+
(1-\beta)\bar{D}_{k,j},
\label{eq:band-ema}
\end{equation}
where $\beta\in(0,1]$ determines the weight assigned to the most recent completed episode.

When DFS selects a target within band $\mathcal{B}_k$, ADM sets $H=\max(\bar{D}_k,1)$, where $\bar{D}_k$ denotes the latest persistence estimate for that band.
ADM uses $H$ as a heuristic horizon, measured in iterations, to evaluate the potential energy gain of a downshift.

\subsubsection{Frequency Adjustment Decision}
\label{sec:frequency-adjustment-decision}

ADM prioritizes upshifts to protect latency and evaluates downshifts to determine whether their potential energy savings justify the transition cost.
Upshifts bypass this assessment, while downshifts need to pass a probabilistic benefit--cost check.

Let $f_{\mathrm{cur}}$ denote the current frequency and $f_{\mathrm{tgt}}$ denote the frequency selected by DFS using Eq.~\eqref{eq:ftgt}.
When $f_{\mathrm{tgt}}>f_{\mathrm{cur}}$, ADM directly initiates the upshift to ensure latency SLOs are met.
When $f_{\mathrm{tgt}}=f_{\mathrm{cur}}$, ADM maintains the frequency.

For a downshift where $f_{\mathrm{tgt}}<f_{\mathrm{cur}}$, ADM first estimates the energy saving per upcoming iteration:
\begin{equation}
\Delta E
=
E(f_{\mathrm{cur}})
-
E(f_{\mathrm{tgt}}).
\label{eq:delta-e}
\end{equation}
Using the estimated horizon $H$ from the previous subsection, the potential energy gain is $E_{\mathrm{gain}}=H\Delta E$.

The corresponding reduction in busy power is
\begin{equation}
\Delta P
=
\max\!\left(
P_{\mathrm{busy}}(f_{\mathrm{cur}})
-
P_{\mathrm{busy}}(f_{\mathrm{tgt}}),
0
\right).
\label{eq:delta-p}
\end{equation}
Let $T_{\mathrm{trans}}$ denote the frequency transition effective time.
The associated energy opportunity cost is approximated as
$E_{\mathrm{cost}}\approx\Delta P\,T_{\mathrm{trans}}$.
Equating this cost with the potential energy gain yields the break-even transition time:
\begin{equation}
T_{\mathrm{BE}}=
\begin{cases}
H\Delta E/\Delta P, & \Delta P>0,\\
+\infty, & \Delta P=0.
\end{cases}
\end{equation}
Under this cost approximation, $T_{\mathrm{BE}}$ is the maximum transition duration for which the transition opportunity cost does not exceed the potential energy gain.

ADM permits a downshift only when the per-iteration energy saving is positive and the transition is sufficiently likely to complete within the break-even duration:
\begin{equation}
\Delta E>0,\qquad
\widehat{F}(T_{\mathrm{BE}})\geq\eta.
\label{eq:adjustment-gate}
\end{equation}
Here, $\widehat{F}(t)$ defined in Eq. (\ref{eq:empirical-cdf}) denotes the empirical CDF of frequency transition effective time, and $\eta$ specifies the minimum required probability that the transition completes within $T_{\mathrm{BE}}$.

\subsection{Support for Colocated and Disaggregated Serving}
\label{sec:pd-disaggregation}

\adaptshift{} supports both colocated serving and prefill--decode (PD) disaggregation.
In colocated serving, prefill and decode share the same serving instance~\cite{yu2022orca,kwon2023vllm}.
The scheduler can produce separate prefill and decode iterations or combine both types of work in a mixed iteration~\cite{agrawal2024sarathi,zheng2024sglang,holmes2024deepspeedfastgen}.
\adaptshift{} runs a controller for each instance and adjusts its GPU frequency according to the upcoming iteration's workload and latency budget.

In PD disaggregation, \adaptshift{} runs an independent frequency controller for each prefill and decode instance.
Prefill instances process input tokens and transfer the resulting KV cache to decode instances, which generate output tokens~\cite{zhong2024distserve,patel2024splitwise,qin2024mooncake,hu2024tetriinfer}.
Each controller applies ILP, DFS, and ADM to its local forward iterations, allowing frequency selection and adjustment to adapt to the workload and latency constraints of each stage.

The two stages differ in their workload composition and latency budgets.
DFS applies a TTFT-derived latency budget to prefill iterations and a TPOT-derived budget to decode iterations, with queue-aware tightening based on the local waiting-queue length, as defined in Eq.~\eqref{eq:slo-budget}.
The prefill budget needs to leave sufficient slack for request queuing, KV-cache transfer, and the first decode iteration, which also contribute to end-to-end TTFT~\cite{zhong2024distserve,patel2024splitwise}.


\section{Implementation}
\label{sec:implementation}

We implement \adaptshift{} as a Python runtime module integrated into SGLang's scheduler~\cite{zheng2024sglang}.

\subsection{Scheduler Integration}
\label{sec:impl-scheduler}

SGLang invokes the controller after selecting the next batch and prior to its forward pass.
This integration covers normal and overlap scheduler loops, alongside separate prefill and decode event loops for PD disaggregation.
Both deployment modes call the same controller entry point with the selected batch and local waiting-queue length.
The hook reads the batch mode to identify prefill, decode, and mixed execution.
Afterward, SGLang processes the batch via its standard execution path.

\subsection{Frequency Actuation}
\label{sec:impl-actuation}

\adaptshift{} sets the frequency via the NVIDIA Management Library (NVML) \texttt{nvmlDeviceSetGpuLockedClocks}~\cite{nvidia2026nvml}.
It tracks the target frequency independently of the NVML-reported value to separate control logic from hardware feedback.
Command completion does not mean that the new frequency affects kernel performance; thus, ADM uses the profiled transition effective times in Section~\ref{sec:effective-time-cdf}.

\subsection{Frequency and Power Profiling}
\label{sec:impl-idle-profiling}

\adaptshift{} uses SM-frequency control ranges of 210--1740~MHz on A40, 690--1830~MHz on H20, and 420--2062~MHz on GB200.
These GPUs support SM-frequency adjustment at 15~MHz granularity.
For busy and locked-idle power profiling, we configure 13, 10, and 14 sampling frequencies on A40, H20, and GB200, respectively (approximately 127~MHz spacing).
The busy-power model in Eq.~(\ref{eq:busy-power}) achieves $R^2=0.99$.
We also evaluate frequency-selection quality on 17 held-out configurations of Qwen3-14B TP4 on H20, covering batch sizes of 1, 2, 4, 8, 16, 32, and 64 and KV lengths of 256, 512, 1,024, and 2,048 tokens. The selected frequencies consume 2.06\% more energy than the measured optimum at the median and 15.46\% more at the 95th percentile. In 88.24\% of cases, their energy consumption remains within 5\% of the optimum.

\subsection{Profiling Transition Effective Times}
\label{sec:impl-transition-profiling}

We profile transition effective times offline by detecting changes in probe execution time, following the measurement principle of Veli\v{c}ka et al.~\cite{velicka2025switching}.
Our Python microbenchmark uses FP16 GEMM (\texttt{torch.mm})~\cite{paszke2019pytorch} as a frequency-sensitive probe and times each execution with \texttt{torch.cuda.Event}~\cite{nvidia2026cudaevents}.
At each tested frequency, the median steady-state execution time defines its baseline.
We exclude frequency pairs with indistinguishable baselines.

During each downshift, we repeatedly execute GEMM while requesting a lower frequency through NVML.
We detect a sustained response when eight consecutive executions reach or exceed the midpoint between the initial and target frequency baselines.
The transition effective time runs from the host-side command trigger to completion of the first execution in that sequence.
This interval includes thread wake-up, command submission, and synchronization overheads.
For each GPU, we pool downshift trials across frequency pairs to construct the empirical CDF used by ADM (Section~\ref{sec:effective-time-cdf}).

\subsection{Controllers across GPU Ranks}
\label{sec:impl-ranks}

In a multi-GPU serving instance, each GPU rank maintains local \adaptshift{} state and commands its own GPU.
These local \adaptshift{} instances implement the instance-level procedure described in Section~\ref{sec:pd-disaggregation}.
For synchronized control, \adaptshift{} coordinates TP ranks through SGLang's TP CPU process group.
After local selection and checks, ranks exchange proposals and select the highest frequency within the supported limits.
Each rank submits a command if the target differs from its current setting, then exchanges command errors before resuming forward execution.
This aligns command targets, though physical GPU transitions may finish at varying times. Consequently, each iteration involves two control exchanges via the CPU process group.
Rank coordination remains internal to the serving instance, keeping prefill and decode instances isolated.

\section{Evaluation}
\label{sec:evaluation}

\subsection{Experimental Setup}
\label{sec:eval-setup}

\begin{table}[t]
\centering
\caption{Evaluation deployments and workloads.}
\label{tab:eval-platforms}
\footnotesize
\setlength{\tabcolsep}{3pt}
\begin{tabular}{lllll}
\toprule
GPU & Model & Parallelism & Azure $r$ & Alibaba $r$ \\
\midrule
A40 & Qwen3-8B, Qwen3-14B & TP2, TP2 & $1/8,1/8$ & $1/4,1/8$ \\
H20 & Qwen3-14B, Qwen3-32B & TP4, TP4 & $1/2,1/4$ & $1,1/3$ \\
GB200 & Qwen3-14B, Qwen3-32B & TP1, TP1 & $1,1/2$ & $1,2/3$ \\
GB200 & DeepSeek-V4-Flash & TP4+EP4 & $1/4$ & -- \\
\bottomrule
\end{tabular}
\end{table}

\textbf{Platforms.}
We evaluate \adaptshift{} with three LLMs in SGLang~\cite{zheng2024sglang} across three GPU platforms (Table~\ref{tab:eval-platforms}).
Qwen3~\cite{qwen2025qwen3} uses BF16; DeepSeek-V4-Flash uses the released FP4/FP8 mixed-precision weights~\cite{deepseek2026v4}.

\textbf{Workloads and SLOs.}
We use two production conversation traces from Microsoft Azure~\cite{azure2023llmtrace} and Alibaba Cloud's Qwen-Bailian platform~\cite{qwenbailiantrace}, replaying the first 2,000 requests from each trace in their original order.
Azure's arrival rate ranges from 0 to 12~QPS, averaging 4.71~QPS, with input and output lengths spanning 2 to 7{,}930 and 12 to 1{,}000 tokens, respectively.
Qwen-Bailian's arrival rate ranges from 0 to 10~QPS, averaging 3.40~QPS, with input and output lengths spanning 66 to 18{,}177 and 2 to 4{,}216 tokens, respectively.
The TPOT SLO is 50~ms; TTFT SLOs are 500~ms for inputs below 512 tokens and 2{,}000~ms otherwise.

To account for differences in serving capacity across GPUs and LLMs, we introduce a deployment-specific load variation factor $r$ (Table~\ref{tab:eval-platforms}).
We replay each trace at $r$ times its original arrival rate, preserving request order and input/output lengths.

\textbf{Baselines.} We compare \adaptshift{} with four baselines.
(1) \emph{Default} uses NVIDIA's default frequency management.
(2) \emph{Fixed} locks SM clocks at 1{,}305/1{,}410/1{,}290~MHz on A40/H20/GB200, respectively.
These frequencies are selected via preliminary sweeps to balance energy and latency, applying uniform settings across LLMs and traces per platform.
(3) \emph{throttLL'eM}~\cite{kakolyris2025throttllem} 
projects future batch size and KV-cache usage, predicts iteration throughput with a trained model, and selects the lowest frequency estimated to satisfy the SLO.
(4) \emph{GreenLLM}~\cite{liu2026greenllm} is a feedback-based frequency-scaling method in SGLang, adjusting GPU frequency based on recent token throughput and latency to reduce energy under SLO constraints.

\textbf{Metrics.}
We evaluate energy efficiency and serving performance using GPU energy savings, SLO violation rates, goodput, and output tokens per joule (TPJ).
We compute total GPU energy by summing changes in NVML's cumulative energy counters~\cite{nvidia2026nvml} across GPUs.
For each trace and deployment, we report the energy ratio $E/E_{\mathrm{Default}}$ and percentage savings $100(1-E/E_{\mathrm{Default}})$ using the matched Default run.
TTFT and TPOT violation rates measure the percentages of requests that violate the respective SLOs.
Goodput (requests/s) counts successful requests satisfying both the TTFT and TPOT SLOs per second~\cite{zhong2024distserve}.

\begin{figure}[t]
\centering
\setlength{\abovecaptionskip}{3pt}
\includegraphics[width=\columnwidth]{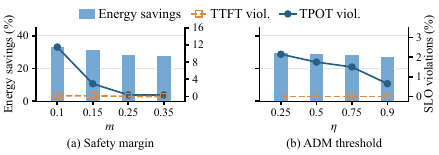}
\caption{Parameter sensitivity on Qwen3-14B TP4 + H20. (a) Safety margin sweep with $\eta=0.75$. (b) ADM threshold sweep with $m=0.25$.}
\label{fig:eval-sensitivity}
\end{figure}

\subsection{Parameter Tuning}
We evaluate the effects of \adaptshift{}'s two parameters, safety margin $m$ and ADM threshold $\eta$, using Qwen3-14B with TP4 on H20 GPUs.
Accordingly, we set $m$=0.25 and $\eta$=0.75 for subsequent experiments.

\textbf{The safety margin $m$ tightens frequency feasibility.}
A larger $m$ reserves more latency budget for prediction error and host overhead, excluding frequencies near the predicted SLO boundary.
As shown in Figure~\ref{fig:eval-sensitivity}(a), increasing $m$ from 0.10 to 0.25 decreases TPOT violations from \evalnum{margin0.1-tpot}\% to \evalnum{margin0.25-tpot}\% and reduces energy savings from \evalnum{margin0.1-saving}\% to \evalnum{margin0.25-saving}\%.
Further tightening to $m=0.35$ leaves both metrics nearly unchanged, indicating diminishing returns at this load.

\textbf{The ADM threshold $\eta$ controls downshift acceptance.}
A larger $\eta$ requires a higher probability for downshifts to complete within the energy break-even time, making ADM more conservative.
In Figure~\ref{fig:eval-sensitivity}(b), raising $\eta$ from 0.25 to 0.90 decreases TPOT violations from \evalnum{eta0.25-tpot}\% to \evalnum{eta0.9-tpot}\% and reduces energy savings from \evalnum{eta0.25-saving}\% to \evalnum{eta0.9-saving}\%.
TTFT violations remain zero across this sweep.

\textbf{Empirical settings reflect workload dynamics.}
For H20, we choose $\gamma=0.2$ in Eq.~\eqref{eq:slo-budget} so that five queued requests halve the iteration budget, prioritizing queue clearance during arrival bursts.
We set $\alpha=0.2$ in Eq.~\eqref{eq:iter-ema} to retain several iterations of timing history, smoothing isolated gaps while following load changes that persist across iterations.
Band statistics update only when an episode ends; we choose $\beta=0.3$ in Eq.~\eqref{eq:band-ema} to give each completed episode more weight than an individual timing observation when updating workload persistence.
Similar energy near the minimum (Observation~1) motivates a nominal $w=100$~MHz as a heuristic grouping width (Section~\ref{sec:frequency-persistence}).
This grouping aims to absorb minor target changes during sustained loads while keeping larger frequency shifts separate.

\begin{figure*}[t]
\centering
\includegraphics[width=\textwidth]{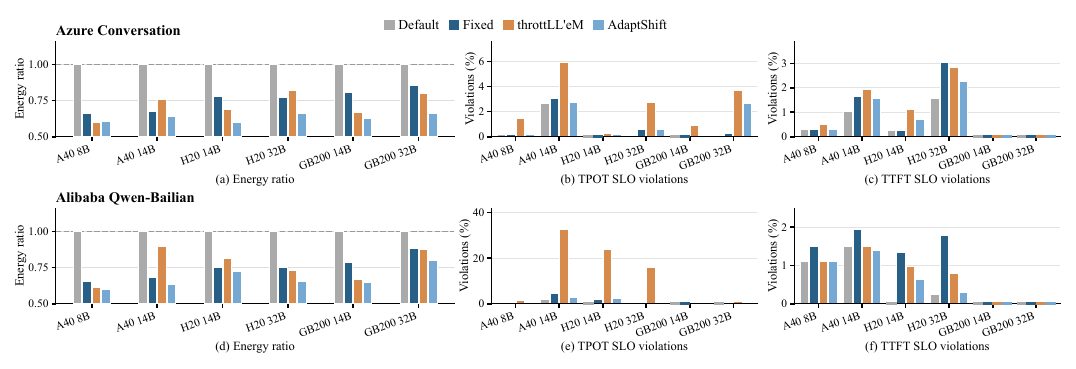}
\caption{Energy ratios and SLO violations on Azure Conversation and Alibaba Qwen-Bailian. All runs replay the first 2{,}000 requests at the deployment-specific load variation factor $r$ in the setup.}
\label{fig:eval-dense}
\end{figure*}
\providecommand{\evalnum}[1]{\ifcsname eval@#1\endcsname\csname eval@#1\endcsname\else\textcolor{blue}{\textbf{XX}}\fi}

\subsection{End-to-End Dense Serving Results}
\label{sec:eval-e2e}

\adaptshift{} reduces GPU energy consumption and lowers TTFT and TPOT SLO violation rates by up to 2.67× and 22.5×, respectively, relative to throttLL’eM across the Azure Conversation and Alibaba Qwen-Bailian traces (Figure~\ref{fig:eval-dense}).
The current estimates also indicate limited goodput loss: \adaptshift{} maintains 96.8–100\% of Default’s goodput across both traces.

\adaptshift{} exploits workload variation to save 20.0\% to 40.2\% GPU energy over Default across both traces.
It also achieves 1.11--2.40$\times$ Fixed's energy savings across the six deployments on each trace.
Default uses NVIDIA's default frequency management, while Fixed uses a fixed GPU frequency throughout the replay.
As batch sizes, context lengths, and prefill/decode mixes change, iterations leave different amounts of latency slack for frequency scaling.
\adaptshift{} selects the energy-optimal frequency among the frequencies that satisfy the iteration's SLO budget, using lower-energy operating points when latency slack permits and higher frequencies when the workload tightens the budget.

Compared with throttLL'eM, \adaptshift{} achieves up to \evalnum{dense-thrott-saving-ratio-max}$\times$ energy savings on the Azure Conversation trace and 3.46$\times$ on the Qwen-Bailian Conversation trace.
Both TTFT and TPOT SLO violation rates remain below 3\% across the two traces.
throttLL'eM~\cite{kakolyris2025throttllem} selects the lowest predicted SLO-feasible decode frequency, whereas \adaptshift{} selects the energy-optimal frequency for both prefill and decode.
Using ILP's latency predictions, DFS compares predicted busy and locked-idle energy across SLO-feasible frequencies (Eq.~\ref{eq:energy}).
ADM accounts for workload persistence and transition opportunity cost to avoid downshifts with insufficient potential energy savings.
It effectively balances energy saving and SLO assurance.

\begin{table}[t]
\centering
\caption{Results of PD disaggregation and EP. TTFT and TPOT columns report SLO violation rates.}
\label{tab:eval-pd-moe}
\footnotesize
\setlength{\tabcolsep}{3pt}
\renewcommand{\arraystretch}{1.08}
\begin{tabular*}{\columnwidth}{@{}l@{\extracolsep{\fill}}rrrr@{}}
\toprule
Method & Savings (\%) & TTFT (\%) & TPOT (\%) & Goodput (req/s) \\
\midrule
\multicolumn{5}{@{}l}{\textbf{Prefill/decode disaggregation}} \\
\addlinespace[2pt]
Default & -- & \evalnum{pd-def-ttft} & \evalnum{pd-def-tpot} & \evalnum{pd-def-goodput} \\
Fixed & \evalnum{pd-fixed-saving} & \evalnum{pd-fixed-ttft} & \evalnum{pd-fixed-tpot} & \evalnum{pd-fixed-goodput} \\
GreenLLM & \evalnum{pd-green-saving} & \evalnum{pd-green-ttft} & \evalnum{pd-green-tpot} & \evalnum{pd-green-goodput} \\
\adaptshift{} & \evalnum{pd-tafs-saving} & \evalnum{pd-tafs-ttft} & \evalnum{pd-tafs-tpot} & \evalnum{pd-tafs-goodput} \\
\midrule
\multicolumn{5}{@{}l}{\textbf{Expert parallelism}} \\
\addlinespace[2pt]
Default & -- & \evalnum{moe-def-ttft} & \evalnum{moe-def-tpot} & \evalnum{moe-def-goodput} \\
Fixed & \evalnum{moe-fixed-saving} & \evalnum{moe-fixed-ttft} & \evalnum{moe-fixed-tpot} & \evalnum{moe-fixed-goodput} \\
throttLL'eM & \evalnum{moe-thrott-saving} & \evalnum{moe-thrott-ttft} & \evalnum{moe-thrott-tpot} & \evalnum{moe-thrott-goodput} \\
\adaptshift{} & \evalnum{moe-tafs-saving} & \evalnum{moe-tafs-ttft} & \evalnum{moe-tafs-tpot} & \evalnum{moe-tafs-goodput} \\
\bottomrule
\end{tabular*}
\end{table}

\subsection{Applicability to PD and EP}
\label{sec:eval-extensions}

\adaptshift{} extends beyond colocated dense-model serving to PD disaggregation and MoE EP.
Table~\ref{tab:eval-pd-moe} reports both experiments using the first 2{,}000 requests from the Azure trace.

The disaggregated deployment serves Qwen3-14B on six H20 GPUs at $r=0.5$, allocating four GPUs to two TP2 prefill replicas and two GPUs to one TP2 decode replica.
The SGLang router directs requests to the prefill replicas, which transfer their KV caches to the decode replica through NIXL~\cite{nvidia2026nixl}.
\adaptshift{} adjusts frequencies independently in the prefill and decode replicas, reducing energy below \emph{Fixed} and \emph{GreenLLM} while maintaining zero TTFT and TPOT violations and nearly unchanged goodput.

The EP deployment serves DeepSeek-V4-Flash with TP4 + EP4 on four GB200 GPUs at $r=0.25$.
DeepEP dispatches tokens to expert ranks and combines their outputs~\cite{deepseek2025deepep}.
Supplementary measurements under this setup quantify actual routing imbalance for each layer and sampled forward as $\rho=\max_g L_g/(\frac{1}{4}\sum_g L_g)$, where $L_g$ counts token--expert assignments received by rank $g$, excluding padding.
The median, p95, and maximum $\rho$ are 1.205, 1.405, and 1.764, respectively.
ILP estimates each rank's expert-token load from cumulative routing counts and uses the slowest rank's predicted expert time (Section~\ref{sec:iteration-level-latency-estimation}).
AdaptShift reduces GPU energy consumption by 25.75\% and 10.29\% relative to Default and Fixed, respectively, while keeping TTFT violations at zero and TPOT violations at 2.57\%.

The SGLang integration of throttLL'eM incurs \evalnum{moe-thrott-tpot}\% TPOT violations.
Its trained throughput model uses batch size, KV-cache occupancy, frequency, and parallelism size, but omits routed expert loads~\cite{kakolyris2025throttllem}.
At the same batch size, different routing distributions change the slowest rank's workload and hence expert-stage latency (Eq.~\ref{eq:moe-ep-stage}).
This omission leaves its frequency selection vulnerable to latency underestimation, resulting in high TPOT violations.


\begin{figure}[t]
\centering
\setlength{\abovecaptionskip}{3pt}
\includegraphics[width=\columnwidth]{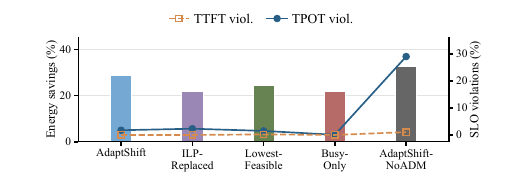}
\caption{Component ablations on 14B TP4 + H20.}
\label{fig:eval-ablation}
\end{figure}

\subsection{Ablations}
\label{sec:eval-dfs-adm}

We compare \adaptshift{} with four variants on Qwen3-14B TP4 + H20 with the first 2{,}000 requests from the Azure trace at $r$=1. All variants share frequency candidates and replay settings, with $m=0.25$ and $\eta=0.75$ when retaining ADM.

(1) ILP-Replaced replaces kernel-based latency composition of ILP with a predictor trained for a specific LLM-GPU configuration, following throttLL'eM~\cite{kakolyris2025throttllem}. We train the replacement predictor on Qwen3-32B TP4 + H20 and apply it to Qwen3-14B TP4 + H20 without retraining to evaluate cross-model prediction accuracy and its impact on energy savings. ILP-Replaced retains DFS, ADM, and the power models.

Each GPU uses a GPU-specific kernel predictor.
The profiling data span 10 frequencies from 690 to 1{,}830~MHz.
The H20 profiling set contains 82{,}790 GEMM, 26{,}880 prefill-attention, 4{,}116 decode-attention, and 1{,}320 all-reduce samples.
For H20, GEMM measurements cover $M=1$ to $16{,}384$, $N=256$ to $152{,}064$, and $K=512$ to $32{,}768$.
Prefill attention covers batch sizes 1 to 8, chunk lengths up to 8{,}192, and prior KV lengths up to 8{,}192 tokens.
Decode attention covers batch sizes up to 64 and KV lengths from 128 to 8{,}192 tokens.
All-reduce covers 10~KiB to 40~MiB.
We randomly split profiling configurations into 80\% for training and 20\% for testing, keeping repeated measurements of each configuration in the same partition.
Kernel profiling takes a total of 1.5 hours on a single GPU.

In Figure~\ref{fig:eval-ablation}, ILP-Replaced saves 21.73\% energy, compared with 28.93\% for \adaptshift{}.
Its TTFT and TPOT violation rates are 0.00\% and 2.31\%, respectively.
On identical held-out 14B configurations and frequencies, ILP-Replaced's prefill, decode, and mixed GPU busy-time MAPEs are 16.20\%, 18.35\%, and 31.70\%, respectively, compared with 5.30\%, 3.42\%, and 7.40\% for ILP.
Prediction errors alter DFS's SLO-feasible candidates and energy estimates, affecting subsequent frequency selection.



 We also evaluate ILP's iteration GPU busy-time prediction accuracy across dense-model and MoE EP deployments, as shown in Table~\ref{tab:eval-iteration-prediction}.
 Mixed iterations contain both prefill and decode work.

(2) \emph{DFS frequency selection.} \emph{Lowest-Feasible} replaces DFS's energy-minimizing selection with the lowest frequency predicted to meet the SLO constraints, retaining ILP and ADM.
This comparison isolates the benefit of energy minimization among SLO-feasible frequencies. 
As shown in Figure~\ref{fig:eval-ablation}, \adaptshift{} improves energy savings over Lowest-Feasible by \evalnum{lf-saving-gap}\%, with TPOT violations of \evalnum{tafs-tpot}\% versus \evalnum{lf-tpot}\%.
DFS balances lower power against longer execution when selecting among feasible frequencies.

(3) \emph{Locked-idle energy modeling.} \emph{Busy-Only} retains ILP, ADM, and DFS's energy-minimizing selection, but includes only busy energy in the objective (Eq.~\ref{eq:energy}).
Comparing it with full \adaptshift{} isolates the contribution of locked-idle energy to frequency selection. Busy-Only lowers energy savings from 28.93\% to 21.67\%.
Its objective omits the frequency-dependent energy between kernel executions.

(4) \emph{ADM adjustment decisions.} \emph{AdaptShift-NoADM} retains ILP and DFS but removes ADM's downshift acceptance check.
Comparing it with full \adaptshift{} isolates the effect of transition-aware adjustment on energy and SLO violations.
As shown in Figure~\ref{fig:eval-ablation}, ADM cuts TPOT violations from 29.05\% to 1.75\% and removes the 1.05\% TTFT violations seen in AdaptShift-NoADM. It raises GPU energy by about 4.4\% over AdaptShift-NoADM. These results show that ADM improves latency protection with a small energy cost by limiting downshifts during delayed frequency changes.

\begin{table}[t]
\centering
\caption{Iteration GPU busy-time prediction MAPE (\%).}
\label{tab:eval-iteration-prediction}
\footnotesize
\setlength{\tabcolsep}{2pt}
\begin{tabular*}{\columnwidth}{@{\extracolsep{\fill}}lllrrr@{}}
\toprule
GPU & Model & Parallelism & Prefill & Decode & Mixed \\
\midrule
A40 & Qwen3-8B & TP2 & \evalnum{iter-a40-8b-prefill-mape} & \evalnum{iter-a40-8b-decode-mape} & \evalnum{iter-a40-8b-mixed-mape} \\
 & Qwen3-14B & TP2 & \evalnum{iter-a40-14b-prefill-mape} & \evalnum{iter-a40-14b-decode-mape} & \evalnum{iter-a40-14b-mixed-mape} \\
\midrule
H20 & Qwen3-14B & TP4 & \evalnum{iter-h20-14b-prefill-mape} & \evalnum{iter-h20-14b-decode-mape} & \evalnum{iter-h20-14b-mixed-mape} \\
 & Qwen3-32B & TP4 & \evalnum{iter-h20-32b-prefill-mape} & \evalnum{iter-h20-32b-decode-mape} & \evalnum{iter-h20-32b-mixed-mape} \\
\midrule
GB200 & Qwen3-14B & TP1 & \evalnum{iter-B200-14b-prefill-mape} & \evalnum{iter-B200-14b-decode-mape} & \evalnum{iter-B200-14b-mixed-mape} \\
 & Qwen3-32B & TP1 & \evalnum{iter-B200-32b-prefill-mape} & \evalnum{iter-B200-32b-decode-mape} & \evalnum{iter-B200-32b-mixed-mape} \\
 & DeepSeek-V4-Flash & TP4+EP4 & \evalnum{iter-B200-flash-prefill-mape} & \evalnum{iter-B200-flash-decode-mape} & \evalnum{iter-B200-flash-mixed-mape} \\
\bottomrule
\end{tabular*}
\end{table}

\subsection{Workload Impacts}
\label{sec:eval-load}
To evaluate workload effects, we compare \adaptshift{} with \emph{Default} on Qwen3-14B TP4 + H20, replaying the first 2{,}000 requests from the Azure trace as load variation factor $r$ increases from 0.25 to 1.5.

\begin{figure}[t]
\centering
\includegraphics[width=\columnwidth]{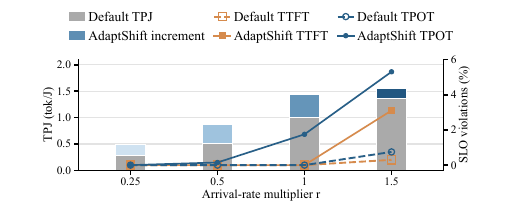}
\caption{Energy efficiency and SLO violations across load variation factor $r$ settings.}
\label{fig:eval-load}
\end{figure}

\adaptshift{} improves energy efficiency across all tested loads, with the largest relative gain at the slowest replay rate, as shown in Figure~\ref{fig:eval-load}.
The gray base of each bar shows \emph{Default}'s efficiency, and the green segment represents \adaptshift{}'s improvement.
Both methods yield higher TPJ as $r$ increases, but \adaptshift{}'s relative gain drops from 73.3\% at $r$=0.25 to 14.2\% at $r$=1.5.
The violation curves show the latency cost of these efficiency gains.
At $r$$\leq$1, \adaptshift{} maintains zero TTFT violations and keeps TPOT violations at or below 1.75\%.
At $r$=1.5, its TTFT and TPOT violations rise to 3.10\% and 5.30\%, respectively.
Consequently, while peak absolute TPJ occurs at high loads, under heavy workloads with strict SLOs, the headroom for adjusting GPU frequency to save energy is limited.

\subsection{Overhead of AdaptShift}
\label{sec:eval-overhead}

We measure \adaptshift{}'s per-iteration control overhead, including prediction and NVML calls, in the main GB200 serving runs (Table~\ref{tab:eval-runtime-overhead}).
The average overhead ranges from \evalnum{runtime-B200-14b-mean-ms} to \evalnum{runtime-B200-flash-mean-ms}~ms per iteration.
The time fraction of \adaptshift{}, computed as mean \adaptshift{} overhead divided by mean iteration time, ranges from \evalnum{runtime-B200-32b-ratio-of-means-pct}\% to \evalnum{runtime-B200-flash-ratio-of-means-pct}\%.
\adaptshift{} executes asynchronously on the CPU, overlapping prediction and frequency control with GPU model computation.
This overlap helps hide control overhead and limits its effect on inference latency.

\begin{table}[t]
\centering
\caption{Overhead of \adaptshift{} on GB200. All times are in ms.}
\label{tab:eval-runtime-overhead}
\footnotesize
\setlength{\tabcolsep}{2pt}
\begin{tabular*}{\columnwidth}{@{\extracolsep{\fill}}lrrrrr@{}}
\toprule
Model & \multicolumn{3}{c}{\adaptshift{} overhead} & \shortstack[r]{Mean iteration\\time} & \shortstack[r]{Time fraction\\(\%)} \\
\cmidrule(lr){2-4}
 & Mean & Median & Max \\
\midrule
Qwen3-14B & \evalnum{runtime-B200-14b-mean-ms} & \evalnum{runtime-B200-14b-p50-ms} & \evalnum{runtime-B200-14b-max-ms} & \evalnum{runtime-B200-14b-iter-mean-ms} & \evalnum{runtime-B200-14b-ratio-of-means-pct} \\
Qwen3-32B & \evalnum{runtime-B200-32b-mean-ms} & \evalnum{runtime-B200-32b-p50-ms} & \evalnum{runtime-B200-32b-max-ms} & \evalnum{runtime-B200-32b-iter-mean-ms} & \evalnum{runtime-B200-32b-ratio-of-means-pct} \\
DeepSeek-V4-Flash & \evalnum{runtime-B200-flash-mean-ms} & \evalnum{runtime-B200-flash-p50-ms} & \evalnum{runtime-B200-flash-max-ms} & \evalnum{runtime-B200-flash-iter-mean-ms} & \evalnum{runtime-B200-flash-ratio-of-means-pct} \\
\bottomrule
\end{tabular*}
\end{table}

\section{Related Work}
\label{sec:related-work}

Existing model-based frequency-selection policies fall into two categories: lowest-feasible selection and energy-model optimization.

\begin{table}[t]
\centering
\caption{Comparison with prior work.}
\label{tab:related-work-comparison}
\scriptsize
\setlength{\tabcolsep}{3pt}
\renewcommand{\arraystretch}{1.1}
\begin{tabularx}{\columnwidth}{@{}l>{\raggedright\arraybackslash}X>{\raggedright\arraybackslash}p{0.14\columnwidth}>{\raggedright\arraybackslash}p{0.17\columnwidth}>{\raggedright\arraybackslash}p{0.12\columnwidth}@{}}
\toprule
\textbf{System} & \textbf{Frequency selection} & \textbf{Transition cost model} & \textbf{Deployment modes} & \textbf{MoE(EP) support} \\
\midrule
throttLL'eM~\cite{kakolyris2025throttllem} & Lowest Feasible & No & Colocated & No \\
EcoInfer~\cite{hu2026ecoinfer} & Static energy floor\newline + Lowest Feasible & No & Colocated & No \\
GreenLLM~\cite{liu2026greenllm} & Queue energy optimization (Prefill); feedback (Decode) & No & PD disaggregation & No \\
DualScale~\cite{basit2026dualscale} & Multi-batch energy optimization (Prefill); Lowest Feasible (Decode) & No & PD disaggregation & No \\
VoltanaLLM~\cite{yu2025voltanallm} & Lowest Feasible & No & PD disaggregation & No \\
\midrule
\adaptshift{} & Load-adaptive energy optimization & Yes & Colocated, PD disaggregation & Yes \\
\bottomrule
\end{tabularx}
\end{table}

(1) Lowest-feasible selection: 
throttLL'eM~\cite{kakolyris2025throttllem}, EcoInfer~\cite{hu2026ecoinfer}, and VoltanaLLM~\cite{yu2025voltanallm} select the lowest predicted SLO-feasible frequency; EcoInfer additionally enforces a static energy floor.
DualScale uses this rule for decode~\cite{basit2026dualscale}.
(2) Energy-model optimization:
GreenLLM's queue energy optimization selects one frequency for all pending prefills within a scheduling window, minimizing active and idle energy under a latency budget~\cite{liu2026greenllm}.
DualScale's multi-batch energy optimization projects the queue into future batches and jointly searches their frequency assignments, allowing each batch a different frequency while satisfying projected TTFT constraints~\cite{basit2026dualscale}.
Additionally, runtime feedback also guides frequency control.
VoltanaLLM~\cite{yu2025voltanallm} updates predictors online, while GreenLLM~\cite{liu2026greenllm} adjusts decode frequency from recent throughput and latency measurements.


\adaptshift{} differs from these systems in frequency selection, transition modeling, and deployment scope.
For frequency selection, \adaptshift{} minimizes the combined busy and locked-idle energy among SLO-feasible frequencies for prefill, decode, and mixed iterations.
It updates these energy estimates as iteration workload and idle time change with serving load.
GreenLLM~\cite{liu2026greenllm} and DualScale~\cite{basit2026dualscale} optimize energy for prefill and use feedback or lowest-feasible selection for decode, respectively.

For adjustment decisions, the compared frequency-control systems acknowledge overhead or delay but report no explicit model of transition effective time~\cite{kakolyris2025throttllem,hu2026ecoinfer,liu2026greenllm,basit2026dualscale,yu2025voltanallm}.
For each downshift, \adaptshift{} combines an empirical distribution of transition effective times with an estimate of how long the selected frequency will remain useful.
It permits the downshift only when potential savings justify transition opportunity cost with the required probability; latency-protecting upshifts take priority.

For deployment scope, \adaptshift{} supports both colocated serving and prefill--decode disaggregation for dense and expert-parallel MoE models.
The compared systems report one of these deployment modes and leave MoE models with EP support unspecified~\cite{kakolyris2025throttllem,hu2026ecoinfer,liu2026greenllm,basit2026dualscale,yu2025voltanallm}.
Table~\ref{tab:related-work-comparison} summarizes these differences in frequency selection, transition modeling, and deployment scope.

\section{Conclusion}

We present \adaptshift{}, an iteration-level frequency scaling framework for energy-efficient LLM serving under latency SLOs.
By combining latency prediction with SLO-constrained energy selection and transition-cost-aware adjustment, \adaptshift{} adapts GPU frequency to changing workloads.
We implement \adaptshift{} in SGLang and conduct extensive experiments to verify its effectiveness.
Future work will extend \adaptshift{} to additional GPU architectures, backends, and parallel configurations, while exploring broader control policies.



\bibliographystyle{IEEEtranS}
\bibliography{refs}

\end{document}